\pdfoutput=1
\documentclass[lettersize,journal]{IEEEtran}

\usepackage[caption=false,font=normalsize,labelfont=sf,textfont=sf]{subfig}
\usepackage{textcomp}
\usepackage{stfloats}

\usepackage{xparse}
\usepackage{amsmath}
\usepackage{commath}
\usepackage{amssymb}
\usepackage{mathtools}
\usepackage{bbm}
\usepackage{physics}
\usepackage{siunitx}
\usepackage{trfsigns}
\usepackage{pifont}
\usepackage{xspace}
\usepackage{etoolbox}
\usepackage{multirow}
\usepackage{csquotes}
\usepackage{graphicx}
\usepackage[shortlabels]{enumitem}  %
\usepackage[nosort]{cite}   %

\usepackage{tablefootnote}

\usepackage{pdfcomment}

\usepackage{placeins}  %

\usepackage{hyperref}
\usepackage[capitalize]{cleveref}
\usepackage [table]{xcolor}
\usepackage{listofitems}
\usepackage{svg}
\usepackage{balance} %
\usepackage{algpseudocode}
\usepackage{algorithm}
\usepackage{booktabs}

\usepackage{url}

\newcommand{\cmark}{\ding{51}}
\newcommand{\xmark}{{--}}

\newcolumntype{H}{>{\setbox0=\hbox\bgroup}c<{\egroup}@{}}   %

\NewDocumentCommand{\tab}{ O{c} O{c} m }{\begin{tabular}[#2]{@{}#1@{}}#3\end{tabular}}   %

\newcommand*{\thl}{\fontseries{b}\selectfont}
\robustify\thl
\robustify\fontseries

\robustify\underline

\usepackage[normalem]{ulem}
\newcommand{\diff}[2]{\if\relax\detokenize{#1}\relax\else\textcolor{red}{\sout{\textcolor{black}{#1}}}\fi\textcolor{green}{#2}}
\newcommand{\changed}[1]{}

\usepackage{ulem}
\colorlet{lightgreen}{green!30}
\colorlet{lightred}{red!30}
\definecolor{forestgreen(web)}{rgb}{0.13, 0.55, 0.13}

\makeatletter
\newcommand\hladd{%
  \bgroup
  \markoverwith{\textcolor{lightgreen}{\rule[-.5ex]{.1pt}{2.5ex}}}%
  \ULon}
\newcommand\hlsout{%
  \bgroup
  \markoverwith{\textcolor{lightred}{\rule[-.5ex]{.1pt}{2.5ex}}\llap{\rule[.5ex]{.3pt}{0.4pt}}}%
  \ULon}
\makeatother
\DeclareRobustCommand{\diff}[2]{\hlsout{#1}\hladd{#2}}

\newcommand{\diffsimple}[2]{\textcolor{red}{#1}{\color{forestgreen(web)}{#2}}}

\newcommand{\tblrowadded}{\smash{\llap{%
\begin{tikzpicture}[baseline=-0.5ex]%
\node[fill=lightgreen,inner xsep=0,scale=0.95](text)at(0,0){\footnotesize new\tikz[baseline=-0.5ex]{\draw[->] (0,0) -- +(1ex,0)}};%
\end{tikzpicture}%
}}}

\DeclareRobustCommand{\diff}[2]{#2}
\DeclareRobustCommand{\diffsimple}[2]{#2}
\renewcommand{\tblrowadded}{}

\newcommand{\vect}[1]{\ensuremath{\boldsymbol{\mathbf{#1}}}}

\renewcommand{\H}{^\mathrm{H}}

\newcommand{\thrDia}{{\tau}}
\newcommand{\thrMasking}{{\xi}}
\newcommand{\thrDistortion}{{\varepsilon}}

\newcommand{\diaProb}{{\tilde{A}_{t, k}}}
\newcommand{\diaEst}[1][t, k]{{\hat{A}_{#1}}}
\newcommand{\diaEstTime}[1][\ell, k]{{\hat{a}_{#1}}}

\newcommand{\mic}{d}
\newcommand{\mics}{D}

\usepackage{pgfplots} %
\newlength\fheight %
\newlength\fwidth %
\usepackage{tikz}
\pgfplotsset{compat=1.9}

\usetikzlibrary{external}

\usetikzlibrary{arrows}
\usetikzlibrary{patterns}
\usetikzlibrary{backgrounds}
\usetikzlibrary{fit}
\usetikzlibrary{positioning}
\usetikzlibrary{shapes.geometric}
\usetikzlibrary{calc}
\usetikzlibrary{shapes.multipart}
\usetikzlibrary{shapes.misc}
\usetikzlibrary{shapes.symbols}
\usetikzlibrary{patterns.meta}
\usetikzlibrary{fadings}

\pgfdeclarelayer{bg}    %
\pgfsetlayers{bg,main}  %

\tikzset{>=stealth}

\tikzstyle{block}=[
    draw,
    text depth=0pt,
    thick, 
    rectangle,
    text centered,
    minimum height=3.4ex,
    fill=white,
    ] %

\tikzstyle{branch}=[{circle,inner sep=0pt,minimum size=0.3em,fill=black}]
\tikzstyle{box}=[rectangle, rounded corners, draw=black, line width=1pt, text width=2cm]

\tikzstyle{arrow}=[{}-{>}, thick]
\tikzstyle{line}=[thick]
\tikzstyle{reverse arrow}=[{<}-{}, thick]

\tikzset{%
	do path picture/.style={%
		path picture={%
			\pgfpointdiff{\pgfpointanchor{path picture bounding box}{south west}}%
			{\pgfpointanchor{path picture bounding box}{north east}}%
			\pgfgetlastxy\x\y%
			\tikzset{x=\x/2,y=\y/2}%
			#1
		}
	},
	sin wave/.style={do path picture={    
			\draw [line cap=round] (-3/4,0)
			sin (-3/8,1/2) cos (0,0) sin (3/8,-1/2) cos (3/4,0);
	}},
	cross/.style={draw, circle, do path picture={    
			\draw [line cap=round] (-2/5,-2/5) -- (2/5,2/5) (-2/5,2/5) -- (2/5,-2/5);
	}},
	plus/.style={draw, circle, do path picture={    
			\draw [line cap=round] (-3/5,0) -- (3/5,0) (0,-3/5) -- (0,3/5);
	}},
	mic/.style={inner sep=0pt, do path picture={
			\draw (0,0) circle (0.9);
			\draw [line cap=round] (-0.9, -0.9) -- (-0.9, 0.9);
	}},
	mux/.style={trapezium, draw}
}

\usepackage[acronym,shortcuts]{glossaries}

\glsdisablehyper
\newacronym{SDR}{SDR}{signal-to-distortion ratio}

\newacronym{CSS}{CSS}{continuous speech separation}
\newacronym{GSS}{GSS}{guided source separation}
\newacronym{PIT}{PIT}{permutation invariant training}
\newacronym{uPIT}{uPIT}{utterance-level permutation invariant training}
\newacronym{MSE}{MSE}{mean squared error}
\newacronym{DFS}{DFS}{depth first search}
\newacronym[]{BLSTM}{BLSTM}{bidirectional long-short-term network}
\newacronym{DPRNN}{DPRNN}{dual-path recurrent neural network}
\newacronym{WER}{WER}{word error rate}
\newacronym{SSE}{SSE}{sum squared error}
\newacronym{DP}{DP}{dynamic programming}
\newacronym{ASDR}{A-SDR}{averaged \gls{SDR}}
\newacronym{SISDR}{SI-SDR}{scale-invariant \gls{SDR}}
\newacronym{SASDR}{SA-SDR}{source-aggregated \gls{SDR}}
\newacronym{SASISDR}{SA-SI-SDR}{source-aggregated scale-invariant \gls{SDR}}
\newacronym{CISDR}{CI-SDR}{convolution invariant \gls{SDR}}
\newacronym{SACISDR}{SA-CI-SDR}{source-aggregated convolution invariant \gls{SDR}}
\newacronym{ASR}{ASR}{automatic speech recognition}
\newacronym{VAD}{VAD}{voice activity detection}
\newacronym{tSDR}{tSDR}{thresholded \gls{SDR}}
\newacronym{logMSE}{log-MSE}{logarithmic mean squared error}
\newacronym{ORCWER}{ORC WER}{optimal reference combination word error rate}
\newacronym{NN}{NN}{neural network}
\newacronym{ORCLEV}{ORC Levenshtein distance}{pptimal reference combination Levenshtein distance}
\newacronym{STFT}{STFT}{short-time Fourier transform}

\newacronym{BCE}{BCE}{binary cross entropy}

\newacronym{cACGMM}{cACGMM}{complex angular central Gaussian mixture model}
\newacronym{cACG}{cACG}{complex angular central Gaussian}

\newacronym{EM}{EM}{expectation maximization}
\newacronym{SMM}{SMM}{spatial mixture model}
\newacronym{MM}{MM}{mixture model}

\newacronym{SAD}{SAD}{speaker activity detection}
\newacronym{IoU}{IoU}{intersection-over-union ratio}
\newacronym{WPE}{WPE}{weighted prediction error}

\newacronym{RIR}{RIR}{room impulse response}
\newacronym{RTF}{RTF}{relative transfer function}

\newacronym{PDF}{PDF}{probability density function}
\newacronym{MVDR}{MVDR}{minimum variance distortionless response}
\newacronym{wMPDR}{wMPDR}{weighted minimum power distortionless response}
\newacronym{ICA}{ICA}{independent component analysis}

\newacronym{cpWER}{cpWER}{concatenated minimum-permutation word error rate}
\newacronym{DIcpWER}{DI-cpWER}{diarization invariant concatenated minimum-permutation word error rate}

\newacronym{DER}{DER}{diarization error rate}
\newacronym{SC}{SC}{spectral clustering}

\newacronym{TSVAD}{TS-VAD}{target-speaker voice activity detection}
\newacronym{TSSEP}{TS-SEP}{target-speaker separation}

\newacronym{SAgWER}{SAg-WER}{speaker agnostic \gls{WER}}

\newacronym{MFCC}{MFCC}{mel frequency cepstral coefficients}

\newacronym{LogMAE}{LogMAE}{logarithmic mean absolute error}

\newacronym{DOA}{DOA}{direction of arrival}

\newacronym{IPD}{IPD}{interchannel phase differences}

\begin{document}

\title{TS-SEP: Joint Diarization and Separation Conditioned on Estimated Speaker Embeddings}
\author{Christoph Boeddeker, Aswin Shanmugam Subramanian, Gordon Wichern,\\
Reinhold Haeb-Umbach, and Jonathan Le Roux

\thanks{
C.\ Boeddeker and R.\ Haeb-Umbach are with Paderborn University, Paderborn, 33098, Germany (e-mail: \{boeddeker,haeb\}@nt.upb.de).}
\thanks{A.\ S.\ Subramanian was with Mitsubishi Electric Research Laboratories (MERL), Cambridge, MA 02139, USA, and is now with Microsoft Corporation, One Microsoft Way, Redmond, WA 98052, USA (e-mail: asubra13@alumni.jh.edu).}
\thanks{
G.\ Wichern and J.\ Le Roux are with Mitsubishi Electric Research Laboratories (MERL), Cambridge, MA 02139, USA (e-mail: \{wichern,leroux\}@merl.com).}
}
\markboth{IEEE/ACM Transactions on Audio, Speech and Language Processing, Vol. XX, 20XX}%
{Boeddeker \MakeLowercase{\textit{et al.}}: TS-SEP: Joint Diarization and Separation Conditioned on Estimated Speaker Embeddings}

\maketitle

\begin{abstract}
Since diarization and source separation of meeting data are closely related tasks, we here propose an approach to perform the two objectives jointly.
It builds upon the target-speaker voice activity detection (TS-VAD) diarization approach, which assumes that initial speaker embeddings are available.
We replace the final combined speaker activity estimation network of \diff{TSVAD}{TS-VAD} with a network that produces speaker activity estimates at a time-frequency resolution. Those act as masks for source extraction, either via masking or via beamforming.
The technique can be applied both for single-channel and multi-channel input and, in both cases, achieves a new state-of-the-art word error rate (WER) on the LibriCSS meeting data recognition task.
We further compute speaker-aware and speaker-agnostic WERs to isolate the contribution of diarization errors to the overall WER performance.
\end{abstract}
\glsresetall


\section{Introduction}

\IEEEPARstart{C}{urrent} research in meeting transcription is not only concerned with  transcribing the audio recordings of meetings into machine-readable text, but also with enriching the transcription with diarization information about ``who spoke when''.
Much of the difficulty of these two tasks  can be attributed to the interaction dynamics of multi-talker conversational speech.
Speakers articulate themselves in an intermittent manner with alternating segments of speech inactivity, single-talker speech, and multi-talker speech.
In particular, overlapping speech, where two or more people are talking at the same time, is known to pose a significant challenge not only to ASR but also to diarization \cite{Watanabe2020CHiME6}.

\glsreset{CSS}

There is no obvious choice whether to start the processing with a diarization or with a separation/enhancement module \cite{Raj2021Meeting}.
An argument in favor of first separating the overlapped speech segments, and at the same time removing other acoustic distortions, and then carrying out diarization, is the fact that the diarization task is much easier when performed on speech without overlap.
This processing order was advocated in \cite{Raj2021Meeting}.
Also, the \gls{CSS} pipeline starts with a separation module \cite{Chen2020LibriCSS}.
Early diarization systems, such as the clustering-based approaches that were predominant in the first editions of the DiHARD diarization challenge, were generally unable to properly cope with overlapping speech, and thus performed poorly on data with a significant amount of concurrent speech \cite{DIHARD_II_2019}. 

On the other hand,  starting the processing with the diarization component can be advantageous, because signal extraction is eased if information about the segment boundaries of speech absence, single-, and multi-talker speech is given. This order of processing has for example been adopted in the CHiME-6 baseline system \cite{Watanabe2020CHiME6}. However, this requires the diarization component to be able to cope with overlapped speech.

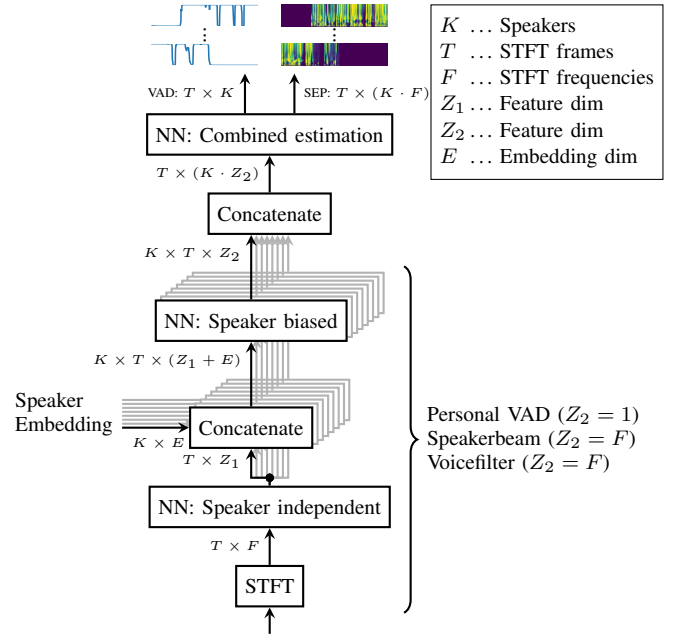
\begin{figure}[!t]
\centering

\begin{tikzpicture}[y=1.5em,x=1.0em,font=\footnotesize]

    \node[block] (stft) at (0, 0) {STFT};
    
    \node[block] (ind) at ($(stft) + (0,2)$) {NN: Speaker independent};
    
	\foreach \i in {7,6,...,0}
	{
    	\ifthenelse{\i=0}{
    		\definecolor{tempcolor}{rgb}{0, 0, 0}
		}{
    		\definecolor{tempcolor}{rgb}{0.7, 0.7, 0.7}
		};
        \node[block, draw=tempcolor, text=tempcolor] (stackEmb-\i) at ($(ind) + (0,2)  + (\i *0.2, \i*0.1) + (-0.7, 0)$) {Concatenate};
        
    	\ifthenelse{\i=0}{
		    \draw[reverse arrow, tempcolor] (stackEmb-\i) -- (stackEmb-\i -| ind.west) -- +(-1.em, 0) node[below right, font=\tiny] {$K \times E$};
    	}{
		    \draw[reverse arrow, tempcolor] (stackEmb-\i) -- (stackEmb-\i -| ind.west) -- +(-1.em, 0);
		};
	    
    	\ifthenelse{\i=0}{
        	\draw[arrow, tempcolor] (ind.north) -- +(0,0.3em) node[branch]{} -| node[above left, font=\tiny] {$T \times Z_1$} (stackEmb-\i.south);
    	}{
        	\draw[arrow, tempcolor] (ind.north) -- +(0,0.3em) node[branch]{} -| (stackEmb-\i.south);
		};
    }
    
    \node[align=left,anchor=east] (permute) at ($(stackEmb-0-|ind.west)!1/2!(stackEmb-7-|ind.west) + (-1em,0)$) {Speaker\\Embedding};

	\foreach \i in {7,6,...,0}
	{
    	\ifthenelse{\i=0}{
    		\definecolor{tempcolor}{rgb}{0, 0, 0}
		}{
    		\definecolor{tempcolor}{rgb}{0.7, 0.7, 0.7}
		};
		
        \node[block, draw=tempcolor, text=tempcolor] (biased-\i) at ($(stackEmb-0 |- stackEmb-7) + (0,2) + (\i *0.2, \i*0.1)$) {NN: Speaker biased};
        
    	\ifthenelse{\i=0}{
	        \draw[arrow, tempcolor] (stackEmb-\i) -- node[left, font=\tiny, pos=0.7] {$K \times T \times (Z_1 + E)$} (biased-\i);
    	}{
	        \draw[arrow, tempcolor] (stackEmb-\i) -- (biased-\i);
		};
    }
    \node[block] (stackSpk) at ($(stft |- biased-7) + (0,2)$) {Concatenate};

    \node[block] (joint) at ($(stackSpk) + (0,2)$) {NN: Combined estimation};
    
	\foreach \i in {7,6,...,0}
	{
    	\ifthenelse{\i=0}{
    		\definecolor{tempcolor}{rgb}{0, 0, 0}
    	    \draw[arrow, tempcolor] (biased-\i) -- node[left, font=\tiny, pos=0.7] {$K \times T \times Z_2$} (stackSpk.south-|biased-\i);
		}{
    		\definecolor{tempcolor}{rgb}{0.7, 0.7, 0.7}
    	    \draw[arrow, tempcolor] (biased-\i) -- (stackSpk.south-|biased-\i);
		};
		
	}
	
	\draw[reverse arrow] (stft.south) -- +(0, -1em);
	\draw[arrow] (stft) -- node[left, font=\tiny] {$T \times F$} (ind);
	\draw[arrow] (stackSpk) -- node[left, font=\tiny] {$T \times (K \cdot Z_2)$} (joint);
	\draw[arrow] ($(joint.north west)!2/5!(joint.north east)$) -- node[left, font=\tiny, align=left] {VAD: $T \times K$} +(0, 1.5em) coordinate (vad);
	
    \node[above left, inner sep=0] (vad1) at ($(vad)+(0.5em, 0.3em)$){\includegraphics[width=5em, height=1em]{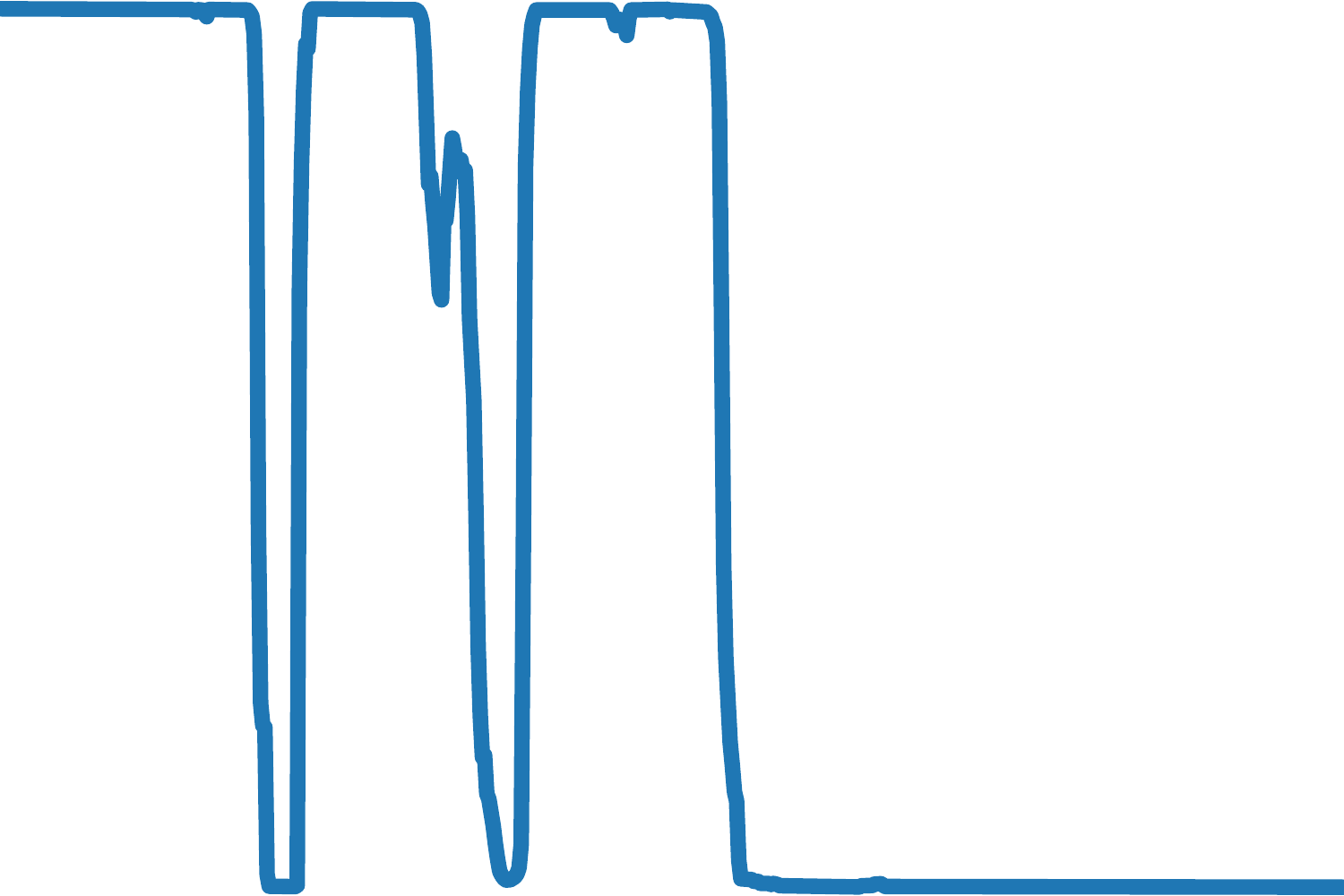}};
    \node[anchor=south, inner sep=0] (vad2) at (vad1.north){\rotatebox{90}{...}};
    \node[anchor=south, inner sep=0] (vad3) at (vad2.north){\includegraphics[width=5em, height=1em]{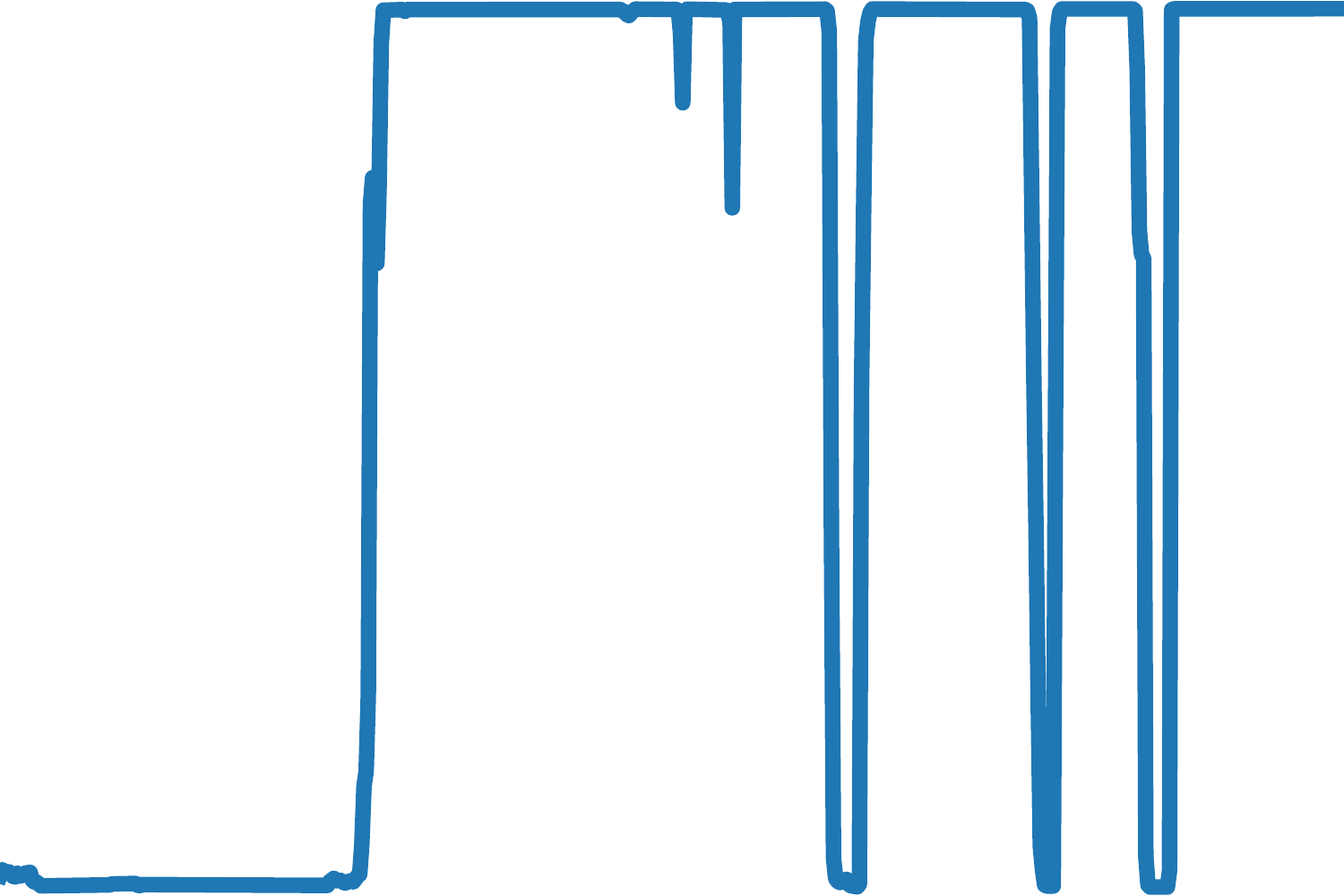}};

	\draw[arrow] ($(joint.north west)!3/5!(joint.north east)$) -- node[right, font=\tiny, align=left] {SEP: $T \times (K \cdot F)$} +(0, 1.5em) coordinate (sep);
	
    \node[above right, inner sep=0] (mask1) at ($(sep)+(-0.5em, 0.3em)$){\includegraphics[width=5em, height=1em]{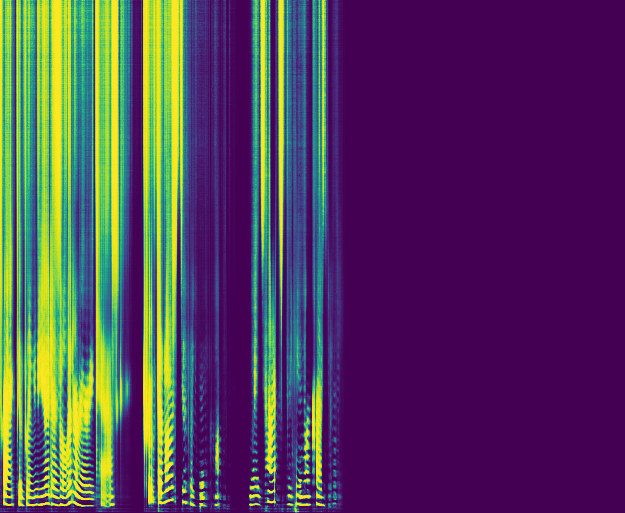}};
    \node[anchor=south, inner sep=0] (mask2) at (mask1.north){\rotatebox{90}{...}};
    \node[anchor=south, inner sep=0] (mask3) at (mask2.north){\includegraphics[width=5em, height=1em]{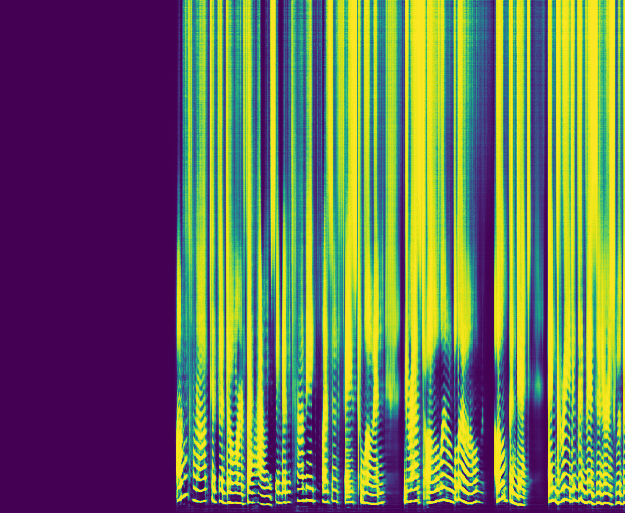}};
	
	\node[align=left, anchor=north west,draw=black, font=\footnotesize] () at ($(ind.east|-mask3.north) + (1.5em, 0em)$) {%
\setlength{\tabcolsep}{0.1em}%
\renewcommand{\arraystretch}{1.1}%
\begin{tabular}{@{}lcl@{}}
	    $K$ & \dots & Speakers\\
	    $T$ & \dots & STFT frames\\
	    $F$ & \dots & STFT frequencies\\
	    $Z_1$ & \dots & Feature dim\\
	    $Z_2$ & \dots & Feature dim\\
	    $E$ & \dots & Embedding dim
    \end{tabular}};
	
	\draw [decorate,
    decoration = {brace, amplitude=0.5em}, thick] ($(biased-7.north east-|ind.south east) + (0.5em,0.2em)$) -- ($(ind.south east|-stft.south) + (0.5em,-0.2em)$) node[pos=0.5,right=0.5em, align=left]{Personal VAD ($Z_2=1$)\\Speakerbeam ($Z_2=F$)\\Voicefilter ($Z_2=F$)};

\end{tikzpicture}

\caption{
TS-VAD and TS-SEP structure for one microphone. Speakers indicated as ``shading''.
}
\label{fig1}
\end{figure}

These different points of view highlight the interdependence of the two tasks, and
they are a clear indication that a joint treatment of diarization and separation/source extraction can be beneficial.
Indeed, subtasks to be solved in either of the two are similar: while diarization is tasked with determining the speakers that are active in each time frame, mask-based source extraction in essence identifies the active speakers for each time-frequency (TF) bin, the difference thus being only the resolution, time vs.\ time-frequency.

Early examples of joint diarization and source separation systems are \glspl{SMM} using time-varying mixture weights \cite{Ito2013permutation, boeddeker22_interspeech}.
There, the estimates of the prior probabilities of the mixture components, after appropriate quantization, give the diarization information about who speaks when, while the posterior probabilities have TF-resolution and can be employed to extract each source present in a mixture, either by masking or by beamforming.
A crucial issue is the initialization of the mixture weights or posterior probabilities.
In the \gls{GSS} framework \cite{Boeddeker2018GSS}, which was part of the baseline system of the CHiME-6 challenge \cite{Watanabe2020CHiME6}, manual annotation of the segment boundaries, and later estimates thereof \cite{Medennikov20CHiME6}, were used to initialize the time-varying mixture weights.
An initialization scheme exploiting the specifics of meeting data, in which most of the time only a single speaker is active, was introduced in \cite{boeddeker22_interspeech}.

However, this approach to joint diarization and separation depends on the availability of multi-channel input, and, due to the iterative nature of the \gls{EM} algorithm, it is computationally demanding.
Furthermore, the \gls{EM} algorithm is, at least in its original formulation, an offline algorithm that is not well suited for processing arbitrarily long meetings.
There exist recursive online variants of the \gls{EM} algorithm, which are nevertheless computationally intense and need careful parameter tuning, in particular in dynamic scenarios \cite{eisenberg21_interspeech}. 

In light of their recent success, both in diarization and source separation, neural networks appear to be promising for developing an integrated solution to the problem of joint diarization and separation/extraction. %
Well-known methods for extracting a target speaker from a mixture include SpeakerBeam \cite{Zmolikova2017SpeakerBeam} and VoiceFilter \cite{Wang2018Voicefilter}.
Both require an enrollment utterance of the target speaker to be extracted.
In the SpeakerBeam approach, the target speaker embedding and the extraction network are trained jointly, while VoiceFilter employs a separately trained speaker embedding computation network.
However, their reliance on the availability of an enrollment sentence somewhat limits their usefulness for some applications.
Furthermore, they have been designed for a typical source separation/extraction scenario, 
where it is a priori known that the target speaker is active most of the time.
Only recently has an extension of the SpeakerBeam system been proposed, to deal with meeting data \cite{Delcroix2021MeetinSpeakerBeam}.

In the field of diarization, end-to-end neural diarization (EEND) has recently made significant inroads, because it can cope with overlapping speech by casting diarization as a multi-label classification problem \cite{Fujita2019EEND,Horiguchi2020EENDEDA,Horiguchi2021EENDEDA,Kinoshita2022EEND}.
EEND processes the input speech in blocks of a few seconds. This introduces a block permutation problem, which can for example be solved by subsequent clustering \cite{Kinoshita_2021}.

\glsreset{TSVAD}
The \gls{TSVAD} diarization system \cite{Medennikov20CHiME6,Medennikov20TS-VAD}, which emerged from the earlier personal VAD approach \cite{Ding2019PersonalVAD}, has demonstrated excellent performance on the very challenging CHiME-6 data set \cite{Watanabe2020CHiME6}.
Assuming that the total number of speakers in a meeting is known, and that embedding vectors representing the speakers are available, \gls{TSVAD} estimates the activity of all speakers simultaneously, as illustrated in \cref{fig1}.
This combined estimation
was shown to be key to high diarization performance.
Rather than relying on enrollment utterances, it estimates speaker profiles from estimated single-speaker regions of the recording to be diarized.
It was later shown that the exact knowledge of the number of speakers is unnecessary, as long as a maximum number of speakers potentially present can be given \cite{He21TS-VAD}, and the attention approach of \cite{Wang2022TSVADTransformer} could do away even with this requirement.

However, \gls{TSVAD} is unable to deliver TF-resolution activity, which is highly desirable to be able to extract the individual speakers' signals from a mixture of overlapping speech.
We here propose a joint approach to diarization and separation that is an extension of the \gls{TSVAD} architecture:
its last layer is replicated $F$ times, where $F$ is the number of frequency bins, see \cref{fig1}. With this modification, the system is now able to produce time-frequency masks.
We call this approach \gls{TSSEP}.

While this may appear to be a small change to the original \gls{TSVAD} architecture, the impact is profound: it results in a joint diarization and separation system that is trained under a common objective function.

Such a joint diarization and separation with \glspl{NN}, where all speakers are simultaneously considered, has so far not been done.
Existing works either do not do the combined estimation of all speakers for separation \cite{Delcroix2021MeetinSpeakerBeam}, which was a key for the success of \gls{TSVAD}, or do experiments on short recordings \cite{Xiao2019SpkExtAtten, Zeghidour2021wavesplit, Maiti2023eend} where diarization is not necessary and hence it is not clear how the approaches would work on long data with many speakers.

In the following, we first introduce the signal model and notations in \cref{sec:signal_model}.
Then, in \cref{sec:tsvad}, we describe the \gls{TSVAD} architecture, from which we derive the proposed \gls{TSSEP} in \cref{sec:tsvad2tssep}.
\Cref{sec:exp} presents a detailed experimental evaluation.
Notably, we achieve a new state-of-the-art WER on the LibriCSS data set.

\section{Signal Model}
\label{sec:signal_model}
We assume that the observation $\vect{y}_{\ell}$ is the sum of $K$ speaker signals $\vect{x}_{\ell, k}$ and a noise signal $\vect{n}_{\ell}$:
\begin{align}
    \vect{y}_{\ell} &= \sum_{k} \vect{x}_{\ell, k} + \vect{n}_{\ell} \in \mathbb{R}^\mics, \label{eq:firstModel}
\end{align}
where $\ell$ is the sample index, and $\vect{y}_{\ell}$, $\vect{x}_{\ell, k}$, and $\vect{n}_{\ell}$ are vectors of dimension $\mics$, the total number of microphones.
We use $\mic$ as microphone index, e.g., $y_{\ell,\mic}$ is the $\mic$'s entrix of $\vect{y}_{\ell}$.
Note that we will also cover the single-channel case, where $\mics=1$.
Since we consider a long audio recording scenario, it is natural to assume that speaker signals
$\vect{x}_{\ell, k}$ have active and inactive time intervals.
We make this explicit by introducing a speaker activity variable $a_{\ell, k} \in \{0, 1\}$:
\begin{align}
    \vect{y}_{\ell} &= \sum_{k} a_{\ell, k} \vect{x}_{\ell, k} + \vect{n}_{\ell}. \label{eq:secModel}
\end{align}
Note that $a_{\ell, k}$ has no effect on the equation, because $\vect{x}_{\ell, k}$ is zero when $a_{\ell, k}$ is zero.
Going to the \gls{STFT} domain, we have
\begin{align}
    \vect{Y}_{t, f} &= \sum_{k} A_{t, k} \vect{X}_{t, f, k} + \vect{N}_{t, f}, %
\end{align}
where $t \in \{1, \ldots ,T\}$ and $f \in \{ 1, \ldots , F\}$ are the time frame and frequency bin indices, and $T$ and $F$ are the total number of time frames and frequency bins, respectively, while $\vect{Y}_{t, f}$, $\vect{X}_{t, f, k}$, and $\vect{N}_{t, f}$ are the STFTs of $\vect{y}_{\ell}$,  $\vect{x}_{\ell, k}$, and  $\vect{n}_{\ell}$, respectively. Further, $A_{t, k} \in \{0, 1\}$ is the activity of the $k$-th speaker at frame $t$, which is derived from $a_{\ell, k}$ by temporally quantizing to time frame resolution.

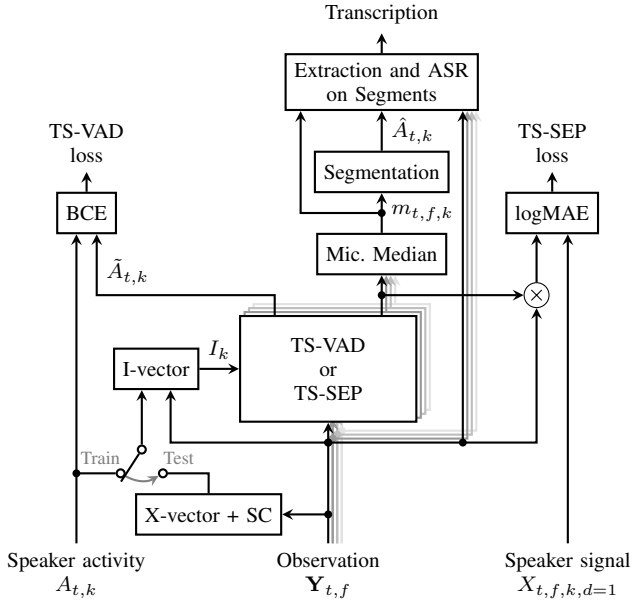
\begin{figure}[!t]
\centering

\begin{tikzpicture}[y=1.5em,x=1.0em,
font=\footnotesize
]

    \node[block, align=center, text depth=, minimum height=4em, text width=6em] (nn) at (0, 0) {TS-VAD \\ or \\ TS-SEP};
    \coordinate (nn-out1) at ($(nn.north west)!1/5!(nn.north east)$);
    \coordinate (nn-out2) at ($(nn.north west)!4/5!(nn.north east)$);
    \coordinate (obs-cross) at ($(nn.south)+(0,-0.5)$);
    
    \node[block, anchor=east] (ivec) at 
    ($(nn.west) + (-1.5, 0)$)
    {I-vector};
    \coordinate (ivec-in1) at ($(ivec.south west)!1/3!(ivec.south east)$);
    \coordinate (ivec-in2) at ($(ivec.south west)!2/3!(ivec.south east)$);

    \node[draw, thick, circle, inner sep=0pt,minimum size=0.3em] (switch-in) at ($(ivec-in1)+(0, -1.5)$) {};
    \node[draw, thick, circle, inner sep=0pt,minimum size=0.3em] (switch-out1) at ($(switch-in)+(-0.8, -0.6)$) {};
    \node[draw, thick, circle, inner sep=0pt,minimum size=0.3em] (switch-out2) at ($(switch-in)+(0.8, -0.6)$) {};
    \draw[arrow, gray] (switch-out1) to[out=-30,in=-150] (switch-out2);
    \draw[thick, shorten >=-0.2em] (switch-in) -- (switch-out1.310);
    
    \node[block, anchor=north west] (sc) at ($(ivec.south|-switch-out1) + (-0.8, -0.5)$) {X-vector + SC};
    \coordinate[] (annot) at ($(ivec.west|-sc.south) + (-1, 0)$);

    \node[block, anchor=south] (median) at ($(nn-out2) + (0, 1)$) {Mic.\ Median};
    
    \node[block, anchor=south] (seg) at ($(median.north) + (0, 1)$) {Segmentation};
    
    \node[block, anchor=south, align=center] (ex) at ($(seg.north) + (0, 1)$) {Extraction and ASR\\on Segments};

    \coordinate (ex-in1) at ($(ex.south-|seg.west) + (-0.5,0)$);
    \coordinate (ex-in3) at ($(ex.south-|seg.east) + (0.5,0)$);
    
    \coordinate (y-loss) at ($(seg)!0.5!(median)$);
    
    \node[block] (bce) at
    ($(y-loss-|annot)$) 
    {BCE};
    \coordinate (bce-in1) at ($(bce.south west)!1/3!(bce.south east)$);
    \coordinate (bce-in2) at ($(bce.south west)!2/3!(bce.south east)$);
    
    \node[block, anchor=west] (logmae) at ($(ex.east|-y-loss) + (1, 0)$) {logMAE};
    \coordinate (logmae-in1) at ($(logmae.south west)!1/3!(logmae.south east)$);
    \coordinate (logmae-in2) at ($(logmae.south west)!2/3!(logmae.south east)$);
    
    \node[cross, anchor=center] (times) at ($(logmae-in1|-nn.north) + (0, 0.5)$) {};
    
    \draw[arrow] (ivec) -- node[above]{$I_k$} (nn.west);

    \node[above, align=center] (obs) at ($(sc.south east-|nn) + (0, -1.8)$) {Observation\\$\vect{Y}_{t, f}$};
    \draw[arrow] (obs) -- (obs|-obs-cross) -| (times);
    \draw[arrow] ($(sc-|obs)$) node[branch]{} -- (sc);
    \draw[arrow] ($(obs-cross-|obs)$) node[branch]{} -| (ivec-in2);
    \draw[arrow] ($(nn.west|-obs-cross)!1/2!(nn.east|-obs-cross)$) node[branch]{} -- ($(nn.south west)!1/2!(nn.south east)$);
    
    \node[anchor=south,align=center] (oracleAct) at (bce-in1|-obs.south) {Speaker activity\\$A_{t,k}$};
    \draw[arrow] (oracleAct) -- (bce-in1);
    
    \draw[arrow] (nn-out1) |- ($(nn-out1-|bce-in2) + (0, 0.5)$) node[above right]{$\diaProb$} -- (bce-in2);
    
    \draw[arrow] (nn-out2|-times) node[branch]{} -- (times);
    
    \draw[arrow] (times) -- (logmae-in1);
    
    \node[anchor=south,align=center] (spkSignal) at (obs.south-|logmae-in2) {Speaker signal\\$X_{t, f, k, \mic=1}$};
    \draw[arrow] (spkSignal) -- (logmae-in2);

    \draw[arrow] (nn-out2) -- (median.south);
    \draw[arrow] ($(median.north)!0.5!(seg.south)$) node[branch]{} -| (ex-in1);
    
    \draw[arrow] (median) -- node[right]{$m_{t,f,k}$} (seg);
    \draw[arrow] (seg) -- node[right]{$\diaEst$} (ex);

    \draw[arrow] ($(ex-in3|-obs-cross)$) node[branch]{} -- (ex-in3);
    
    \draw[line] (sc) -- ($(switch-out2-|sc)$) -- (switch-out2);
    
    \draw[line] ($(switch-out1-|bce-in1)$) node[branch]{} -- (switch-out1);
    \draw[arrow] (switch-in) -- (ivec-in1);
    \node[gray, above left, inner xsep=0, font=\scriptsize] () at (switch-out1) {Train};
    \node[gray, above right, inner xsep=0, font=\scriptsize] () at (switch-out2) {Test};
    
    \draw[arrow] (bce.north) -- +(0,0.5) node[above, align=center]{TS-VAD\\loss};
    
    \draw[arrow] (logmae.north) -- +(0,0.5) node[above, align=center]{TS-SEP\\loss};
    
    \draw[arrow] (ex.north) -- +(0,0.5) node[above, align=center]{Transcription};

    \begin{pgfonlayer}{bg}    %
    	\foreach \i in {6,5,...,1}
    	{
    	    \pgfmathsetmacro\k{0.6+(\i-1)/7}
        	\definecolor{tempcolor}{rgb}{\k,\k,\k}
        	\begin{scope}[shift={(\i *0.17, \i*0.1)}]
                \node[block, draw=tempcolor, align=center, text depth=, minimum height=4em, text width=6em] (nn-\i) at (0, 0) {TS-VAD \\ or \\ TS-SEP};
                \draw[arrow,tempcolor] (obs.north-|nn-\i) -- (nn-\i);
                \draw[arrow,tempcolor] (obs.north-|nn-\i) -- (nn-\i);
                
                \coordinate (nn-out2-\i) at ($(nn-\i.north west)!4/5!(nn-\i.north east)$);
                \draw[arrow,tempcolor] (nn-out2-\i) -- (nn-out2-\i|-median.south);
                
                \draw[arrow,tempcolor] ($(obs-cross)+(nn-\i)-(nn)$) node[branch,tempcolor]{} -| ($(ex-in3)+(nn-\i|-nn)-(nn)$);
                
            \end{scope}
    	}
    \end{pgfonlayer}

\end{tikzpicture}

\caption{
    System overview: Pretraining with TS-VAD loss, training with TS-SEP loss and predicting transcriptions at inference. Microphones indicated as ``shading''.
}
\label{fig:tssep}
\end{figure}

\section{TS-VAD}
\label{sec:tsvad}
\glsreset{TSVAD}
In the CHiME-6 challenge, \gls{TSVAD} \cite{Medennikov20CHiME6,Medennikov20TS-VAD} achieved impressive results for the diarization task, outperforming other approaches by a large margin.
The basic idea is similar to personal VAD \cite{Ding2019PersonalVAD}:
an \gls{NN} is trained to predict the speech activity $A_{t, k}$ of a target speaker $k$ from a mixture $\vect{Y}_{t, f}$, given an embedding $I_k$ representing the speaker, as illustrated in \cref{fig1}.

\Gls{TSVAD} has two main differences to prior work: first, the speaker embeddings are estimated from the recording instead of utilizing  enrollment utterances, as illustrated in \cref{fig:tssep}, and second, a combination layer estimates the activities of all speakers in a segment simultaneously, given an embedding for each speaker, as illustrated in \cref{fig1}.

\subsection{Speaker embedding estimation}

\gls{TSVAD} relies on the availability of initial diarization information, such as from manual annotation or from an embedding extraction system (e.g., X-vector \cite{Snyder2018xVector}) followed by a clustering approach (e.g., spectral clustering \cite{Park2019SpectralClustering}).
Speaker embedding vectors (e.g., I-vectors \cite{Dehak2010IVector}) are then computed from those segments where only a single speaker is active:
\begin{align}
    I_k = \mathrm{Emb}\Big\{y_{\ell,\mic=1}, \forall ~ \ell \text{ s.t. } \diaEstTime = 1 ~\mathrm{ \& }~ \smash{\sum_{\tilde{k}\neq k}} \diaEstTime[\ell, \tilde{k} ]=0 \Big\} \in \mathbb{R}^E.
\end{align}
Here, $\mathrm{Emb}\{\cdot\}$ symbolizes the computation of the embedding vector from segments of speech from the first microphone where a single speaker is active.
Furthermore,  $\diaEstTime$ is an estimate for $a_{\ell, k}$.

While this approach does not require an enrollment utterance, it makes \gls{TSVAD} dependent on another, initial diarization system. Thus,  \gls{TSVAD} can be viewed as a refinement system:
given the estimate of a first diarization, \gls{TSVAD} is applied to estimate a better diarization.
In \cite{Medennikov20TS-VAD}, it was reported that I-vector embeddings are \diff{preferred over}{superior to} X-vectors for \gls{TSVAD}, which is in contrast to other publications that suggest a superiority of X-vectors over I-vectors for diarization tasks \cite{Snyder2018xVector}.
\diff{}{We follow\mbox{\cite{Medennikov20TS-VAD}} and use I-vectors.}

\subsection{TS-VAD architecture}

The \gls{TSVAD} network consists of three components, with stacking operations between them, as shown in \cref{fig1}.
First, the logarithmic spectrogram and the logarithmic mel filterbank features of one microphone of the mixture are stacked and encoded by a few speaker-independent layers, which can be viewed as feature extraction layers, into matrices with dimension $\mathbb{R}^{T \times Z_1}$, where  $Z_1$ is the size of a single framewise representation.

Next, for each speaker $k$, the framewise representation is concatenated with the speaker embedding of that speaker, resulting in an input to the second network component of dimension $\mathbb{R}^{T \times (Z_1+E)}$.
This second network component processes each speaker independently, nevertheless the \gls{NN} layer parameters are shared between the speakers. 
As a consequence, a discrimination of the speakers can only be achieved through the speaker embeddings, not through  the \gls{NN} layer parameters.
For each speaker, the output of the second network component has a dimension of $\mathbb{R}^{T \times Z_2}$, with $Z_2$ denoting the size per frame.
Until this point, the \gls{NN} design matches that of personal VAD \cite{Ding2019PersonalVAD}.

Before the last \gls{NN} layers, the hidden features of all speakers are concatenated to obtain one large feature matrix of dimension $\mathbb{R}^{T \times (K \cdot Z_2)}$.
The final layers produce an output of dimension $\mathbb{R}^{T \times K}$, which, after thresholding and smoothing (see \cref{sec:segmentatuon}), gives the activity estimate 
$\diaEst$ 
for all speakers $k$ and frames $t$ simultaneously.
This combined estimation makes it easier to  distinguish between similar speakers.
In fact, it was shown in \cite{Medennikov20TS-VAD} that the combined estimation layers at the end were instrumental in obtaining good performance.

From this description, it is obvious that \gls{TSVAD} assumes knowledge of the total number $K$ of speakers in the meeting to be diarized, because $K$ defines the dimensionality of the network output.
This constraint can be relaxed by incorporating an attention mechanism as was shown in \cite{Wang2022TSVADTransformer},  
for the case of  fully overlapped speech separation.
Nevertheless, we keep the original \gls{TSVAD} stacking, only increasing the number of speakers from 4 in  \cite{Medennikov20TS-VAD} to 8. We do so  first because  an upper bound of 8 speakers is already large for many meetings, and second to allow for a  better comparison with \cite{He21TS-VAD}, where this stacking is also used.

\subsection{From single-channel to multi-channel}
\label{sec:multichannel}
When multiple channels are available, there are different approaches to utilize them,
e.g., via spatial features like \gls{IPD} \cite{Wang2018IPD}, for intermediate signal reduction such as in an attention layer \cite{Medennikov20TS-VAD}, and output signal reduction \cite{Heymann_Drude_Haeb-Umbach_2016}.
Each of them has different advantages and disadvantages.
We use here an output signal reduction: this means the \gls{NN} is applied independently to each channel and the output is reduced to a single estimate with a median operation.
Incorporating multiple channels in this way allowed us to train the \gls{NN} with a single channel and use all channels at test time.
Besides this, this makes our method agnostic to the array geometry and number of microphones, and the presence of
moving speakers, as in CHiME-6 \cite{Watanabe2020CHiME6}, becomes less critical.
Finally, this approach was shown to be robust to a mismatch between simulated training data and real test data \cite{Heymann_Drude_Haeb-Umbach_2016}.

For ease of notation, we ignore microphone indices for the \gls{NN} input and output in the following section.
At training time, we use only a reference microphone, and at test time, the \gls{NN} is applied independently to each microphone and then the median is used.

\section{From TS-VAD to TS-SEP}
\label{sec:tsvad2tssep}
\glsreset{TSSEP}

We are now going to describe our modifications to the \gls{TSVAD} architecture in order to do joint diarization and separation.
The key difference between diarization and (mask-based) separation is that diarization output has time resolution, while separation requires time-frequency resolution.

\subsection{From Frame to Time-Frequency Resolution}

Starting from the \gls{TSVAD} system, the output size of the last layer is changed from $K$ speakers to $K$ speakers times $F$ frequency bins:
$\mathbb{R}^{T \times K} \rightarrow \mathbb{R}^{T \times (F\cdot K)}$, where $\rightarrow$ denotes the
``repeat''
operation,
\diff{see \mbox{\cref{fig1}}.
}{which is done by repeating the weights of the last layer.}
Rearranging the output from $\mathbb{R}^{T \times (F\cdot K)}$ to $\mathbb{R}^{T \times F \times K}$,
we obtain a $(T \times F)$-dimensional spectro-temporal output for each speaker $k$.
With a sigmoid nonlinearity, which ensures that values are in $[0,1]$, this can be interpreted as a mask $m_{t,f,k}, \forall t,f,k$, that can be used for source extraction, for example via masking:
\begin{align}
    \label{eq:masking}
    \hat{X}_{t,f,k} = m_{t,f,k}Y_{t,f,d=1},
\end{align}
where $d=1$ is the reference microphone.
We denote this system as \gls{TSSEP}.

\diff{\mbox{\Cref{fig:tssep}}}{The sketches at the top of \mbox{\cref{fig1}} illustrate the different outputs of \gls{TSVAD} and \gls{TSSEP} and \mbox{\cref{fig:tssep}}}
gives an overview of \gls{TSSEP}, whose components are described in the following.

\subsection{Training Schedule and Objective}
As \gls{TSSEP} emerged from \gls{TSVAD}, an obvious choice for training is a two-stage training schedule: in the first, pretraining stage, an ordinary \gls{TSVAD} system is trained, until it starts to distinguish between speakers;
then, the last layer is copied $F$ times to obtain the desired time-frequency resolution;
the second training stage is then the training of the \gls{TSSEP} system, initialized with the parameters of the \gls{TSVAD} system.

For the \gls{TSVAD} pretraining, we used the training loss  proposed in the original publication \cite{Medennikov20TS-VAD}, which is the sum of the \gls{BCE} losses between estimated and ground-truth activities for all speakers.

There are several choices for the training objective of the \gls{TSSEP} system.
We opted for a time-domain reconstruction loss, since it implicitly accounts for phase information (see \cite{Heitkaemper2020Demystifying,Cord2021ToReverb} for a discussion of time- vs.\@ frequency-domain reconstruction losses).
To be specific, a time-domain signal reconstruction loss is computed by applying an inverse \gls{STFT} to  $\hat{{X}}_{t,f,k}$ to obtain $\hat{{x}}_{\ell, k}$ and  measuring the \gls{LogMAE} from the ground truth ${x}_{\ell, k}$: 
\begin{align}
    \mathcal{L} = \log_{10} \frac{1}{L} \sum_k \sum_\ell \left| \hat{{x}}_{\ell, k} - {x}_{\ell, k} \right|.
\end{align}
Clearly, other loss functions, such as the MAE, \gls{MSE}, or \gls{SDR} can also be employed.
However, it is important to note that 
for reconstruction losses that contain a log operation, the sum across the speakers should be performed before applying the log operation.
Otherwise, the loss is undefined and the training can become unstable if a speaker is completely silent \cite{VonNeuman2022SA-SDR}.

We also experimented with training the \gls{TSSEP} system from scratch.
However, this turned out to be far more sensitive to the choice of training loss.
We found that only a modified binary cross entropy loss, where we artificially gave a higher weight to loss contributions corresponding to ``target present'', was able to learn from scratch, while other choices did not converge at all\footnote{\diff{}{The networks learned first to predict only zeros and were not able to leave these local optima.}}.  
But even with this loss, the final performance was worse than with the two-stage training\footnote{\diff{}{In some preliminary experiments, we obtained \mbox{\SI{8.72}{\percent}} concatenated minimum-permutation word error rate (cpWER) for two-stage training, while training from scratch led to \mbox{\SI{100}{\percent}} cpWER with the LogMAE loss (because of zero estimates) and \mbox{\SI{12.0}{\percent}} cpWER with the binary cross entropy-based loss.}}.
Here, we therefore only report experiments with the described two-stage training.

\glsreset{cpWER}

\begin{figure}
    \centering
    \input{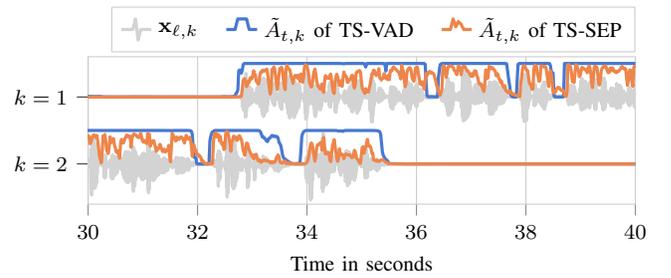}
    \caption{Estimated speaker activity $\diaProb$ for 2 of 8 speakers from the LibriCSS DEV OV40 dataset, shown together with the corresponding clean speaker time signals $\vect{x}_{\ell, k}$. Best viewed in color.}
    \label{fig:activity}
\end{figure}

\begin{figure}
    \centering
    \input{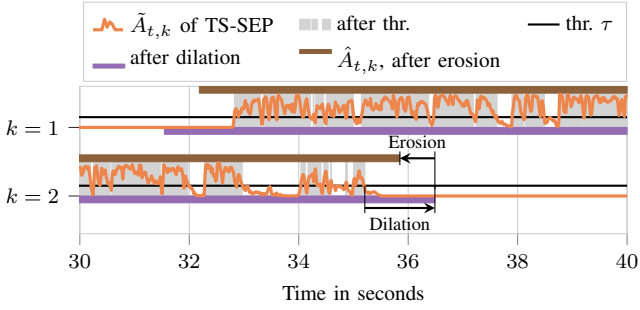}
    \caption{Thresholding and “closing” morphological operation (dilation followed by erosion) on $\diaProb$ to obtain $\diaEst$, for 2 of 8 speakers from the LibriCSS DEV OV40 dataset.
    Best viewed in color.
    }
    \label{fig:closing}
\end{figure}

\subsection{Activity estimation / Segmentation}
\label{sec:segmentatuon}
\gls{TSSEP} outputs a mask with TF resolution. To get an estimate like \gls{TSVAD} with only time resolution, we take the mean across the frequencies:
\begin{align}
    \diaProb &= \frac{1}{F} \sum_f m_{t,f,k}. \label{eq:maskTOvad}
\end{align}
Typically, this estimate has spikes for each word and is close to zero between words, see \cref{fig:activity}.
To fill those gaps, we followed \cite{boeddeker22_interspeech} and used thresholding and the \enquote{closing} morphological operation (see \cref{fig:closing}), known from image processing \cite{Haralick1987imageMorphology}: in a first dilation step, a sliding window is moved over
the signal with a shift of one, and the maximum value inside the
window is taken as the new value for its center sample; in a subsequent erosion step, we
do the same, however with the minimum operation, leading to the final smoothed activity estimate
\begin{align}
    \diaEst &= \operatorname{Erosion}(\operatorname{Dilation}(\delta(\diaProb \geq \thrDia))).
    \label{eq:closing}
\end{align}
Here, $\thrDia$ is the threshold and $\delta(x)$ denotes the Kronecker delta which evaluates to one if $x$ is true and zero otherwise.
When the window for the dilation is larger than the window for erosion, the speech activity is overestimated, i.e., the difference divided by two is added to the beginning and end of the segment, as illustrated in \cref{fig:closing}.
While an overestimation will likely worsen the diarization performance, this may not be the case for \gls{ASR}: it is indeed common to train \gls{ASR} systems with recordings that contain a single utterance plus some silence (or background noise) at the beginning and end of the recording.
We investigate the overestimation later in \cref{sec:ex:seg}.

At activity change points of $\diaEst$, the speakers' signals are cut, leading to segments of constant speaker activity.

This thresholding and \enquote{closing} procedure departs from the median-based smoothing typically used for \gls{TSVAD}. %
When the diarization system is trained to fill gaps of short inactivity (e.g., pauses between words), median-based smoothing is fine.
But in TS-SEP, the system is trained to predict a time-frequency reconstruction mask for each speaker, so typically the mask values are smaller than the VAD values, short inactivity produces zeros, and values are smaller in overlap regions than in non-overlap regions.
The thresholding and closing morphological operations help to fill the gaps.

\subsection{Extraction}

At evaluation time,  an estimate of the signal of speaker $k$ is obtained by  multiplying the mask with the  \gls{STFT} of the input speech, see \cref{eq:masking}.
If multi-channel input is available,  an alternative to mask multiplication for source extraction is to utilize the estimated masks to compute beamformer coefficients \cite{Heymann_Drude_Haeb-Umbach_2016}.
For beamforming, the spatial covariance matrices of the desired signal 
\begin{align}
    \vect{\Phi}_{xx, f, b} &= \frac{1}{|\mathcal{T}_b|} \sum_{t \in \mathcal{T}_b} m_{t,f,k_b} \vect{Y}_{t,f} \vect{Y}_{t,f}\H 
\end{align}
and of the distortion%
\begin{align}
    \vect{\Phi}_{dd, f, b} &= \frac{1}{|\mathcal{T}_b|} \sum_{t \in \mathcal{T}_b} \max\Bigg(\thrDistortion, \sum_{\tilde{k} \neq k_b} m_{t,f, \tilde{k}}\Bigg) \vect{Y}_{t,f} \vect{Y}_{t,f}\H
\end{align}
are first computed.
Here, $b$ denotes the segment index, $\mathcal{T}_b$ \diff{}{is} the set of frame indices that belong to segment $b$, $k_b$ \diff{}{is} the index of the speaker active in segment $b$ who is to be extracted\footnote{The segments are obtained from $\diaEst$, such that each segment $b$ is associated with a single target speaker $k_b$;
when multiple speakers are active at the same time, each speaker gets a different segment index $b$.}, \diff{and}{} $\thrDistortion=0.0001$ \diff{}{is} a small value introduced for stability, \diff{}{and $\sum_{\tilde{k} \neq k_b}$ denotes the summation over all speaker indices, except $k_b$.}

With these covariance matrices, the beamformer coefficients can be computed for example using a \gls{MVDR} beamformer in the formulation of \cite{Souden2009MVDR}:
\begin{align}
    \vect{w}_{f, b} &= \mathrm{MVDR}(\vect{\Phi}_{xx, f, b}, \vect{\Phi}_{dd, f, b}).
\end{align}
Finally, source extraction is performed by applying the beamformer to the input speech:
\begin{align}
    \hat{X}_{t,f,k_b} &= \vect{w}_{f, b}\H \vect{Y}_{t,f}, \quad\forall \: {t \in \mathcal{T}_b}.
 \end{align}
In our experiments, we also combined beamforming with mask multiplication \cite{subramanian2020far}, which led to somewhat better suppression of competing speakers:
\begin{align}
    \hat{X}_{t,f,k_b} &= \vect{w}_{f, b}\H \vect{Y}_{t,f} \max (m_{t,f,k_b}, \thrMasking), \quad\forall \: t \in \mathcal{T}_b,
\end{align}
where $\thrMasking \in [0, 1]$ is a lower bound/threshold for the mask.

\diff{}{The lower bound can be motivated by the observations from\mbox{\cite{iwamoto22_interspeech}}:
pure \gls{NN} based speech enhancement can degrade the \gls{ASR} performance compared to doing nothing;
when adding a scaled version of the observation to the enhanced signal, the \gls{ASR} performance can improve compared to doing nothing.}

\subsection{Guided Source Separation (GSS)}
\label{sec:gss}
An alternative to the immediate use of the mask for extraction is to  fine-tune the mask first.
If multi-channel input is available, this can be achieved by an \gls{SMM}, which utilizes spatial information.
Here, we employ the \gls{GSS}
proposed in \cite{Boeddeker2018GSS}, where the guiding information is given by the diarization information $\diaEst$:
the posterior probability that the $k$-th mixture component is active in a frame $t$ can only be non-zero if $\diaEst$ is one.
The \gls{SMM} is applied to segments $b$ whose boundaries are determined by $\diaEst$,
plus some context.
The \gls{EM} iterations using the guiding information are complemented with a non-guided \gls{EM} step.
The finally estimated class posterior probabilities form the speaker-specific time-frequency masks that are  used for source extraction by beamforming.
For details on \gls{GSS}, see \cite{Boeddeker2018GSS}.

While earlier works initialized the \gls{SMM} by broadcasting \gls{VAD} information $\diaEst$ over the frequency axis that was available either from manual annotation (CHiME-5 \cite{Barker2018CHiME5}) or determined automatically (CHiME-6 Track 2 \cite{Watanabe2020CHiME6}), we here propose to initialize the posterior of the \gls{SMM} with $m_{t, f, k}$, which is only available for \gls{TSSEP}, while voice activity information $\diaEst$ is available for \gls{TSVAD} and \gls{TSSEP}.

\section{Experiments}
\label{sec:exp}
\subsection{Training and Test Datasets}
Diarization and recognition experiments are carried out on the LibriCSS data set, which is a widely used data set for evaluating meeting recognition systems \cite{Chen2020LibriCSS}.
It contains re-recordings of loudspeaker playback of LibriSpeech sentences mixed in a way to reflect a typical meeting scenario. 
The dataset consists of 10 one-hour-long sessions, where each session is subdivided into six 10-minute-long mini sessions that have different overlap ratios, ranging from \num{0} to \SI{40}{\%}.
In each session, a total of eight speakers are active. 
The recording device was a seven-channel circular microphone array.
Following \cite{Raj2021Meeting}, we used the first session as development (DEV) set and the other 9 as evaluation (EVAL) set.

\begin{figure}
    \centering
    \begin{tikzpicture}

\definecolor{darkslategray38}{RGB}{38,38,38}
\definecolor{darkslategray66}{RGB}{66,66,66}
\definecolor{lightgray204}{RGB}{204,204,204}
\definecolor{peru21713894}{RGB}{217,138,94}
\definecolor{steelblue89124191}{RGB}{89,124,191}

\definecolor{royalblue72120208}{RGB}{72,120,208}
\definecolor{coral23813374}{RGB}{238,133,74}

\pgfplotstableread[row sep=\\,col sep=&]{
 Speakers &  NoMixup &  Mixup &  MixupProb \\
        0 &      0.0 &    0.0 &        0.0 \\
        1 &      0.0 &    0.0 &        0.0 \\
        2 &      0.0 &    0.0 &        0.0 \\
        3 &      1.0 &    0.0 &        0.0 \\
        4 &      8.0 &    0.0 &        4.0 \\
        5 &     32.0 &    0.0 &       16.0 \\
        6 &     39.0 &    4.0 &       22.0 \\
        7 &     17.0 &   24.0 &       20.0 \\
        8 &      3.0 &   72.0 &       37.0 \\
    }\mydata

\begin{axis}[
    ybar,
    bar width=.7em,
    width=\columnwidth,
    height=.34\columnwidth,
    legend style={at={(0,1)},
        anchor=north west,legend columns=-1,
        /tikz/every even column/.append style={column sep=0.8em},  %
        },
    xtick=data,
    ymajorgrids,
    tick align=outside,
    tick pos=left,
    nodes near coords,
    nodes near coords align={vertical},
    nodes near coords style={fill=white, inner sep=0.1ex, outer sep=0.3ex, font=\footnotesize},
    xmin=-0.5,xmax=8.5,
    ytick={0,25,...,100}, 
    ymin=0,ymax=100,
    ylabel={\%},
    xlabel={Number of speakers},
    ticklabel style = {font=\footnotesize},
]
        \addlegendimage{ybar,area legend,royalblue72120208, line width=0,fill=royalblue72120208};
        \addlegendimage{ybar,area legend,coral23813374, line width=0, pattern=crosshatch, preaction={fill, coral23813374}, pattern=crosshatch, pattern color=white};
        
        \addplot[royalblue72120208, line width=0, fill=royalblue72120208] table[x=Speakers,y=NoMixup]{\mydata};
        \addplot[coral23813374, line width=0, preaction={fill, coral23813374}, pattern=crosshatch, pattern color=white] table[x=Speakers,y=MixupProb]{\mydata};
        \legend{{$\zeta=0$ (Original)}, $\zeta=0.5$}
    \end{axis}

\end{tikzpicture}
    \caption{Distribution of the number of speakers active in \num{1}-minute chunks of the training data, depending on the probability $\zeta$ to use superposition/mixup ($\zeta=0$ corresponds to the original dataset without superposition).
    }
    \label{fig:hist:speakers}
\end{figure}
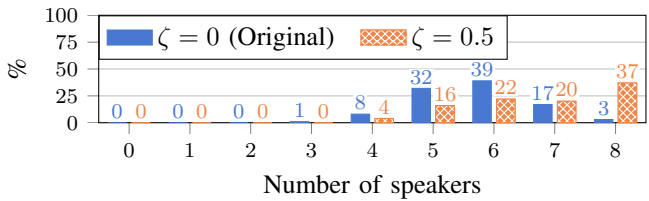

As training data, we employed simulated meeting data created with scripts\footnote{\url{https://github.com/jsalt2020-asrdiar/jsalt2020_simulate}} from the LibriCSS authors.
While each training meeting contains 8 speakers, we trained only on chunks of \num{1} minute. Since within a chunk usually fewer speakers are active, as shown in \cref{fig:hist:speakers}, we superposed segments of speech to artificially increase the number of active speakers, see  \cref{sec:mixup} for a discussion on this.

\subsection{Performance Measures}

 Since this work focuses on meeting transcription enriched with information about who spoke when, we measure both diarization and word error rate performance.

As the diarization performance measure, we employ the widely used \gls{DER}, which is measured with the \enquote{md-eval-22.pl} script from NIST \cite{nist}.
The \gls{DER} is the sum of false alarms, missed hits, and speaker confusion errors divided by the total speaker time.

For the recognition of partially overlapping speech of multiple speakers, there are several \gls{WER} definitions \cite{vonNeumann2022WER}.
Since \gls{TSVAD} groups utterances of the same speaker, we mainly report the \gls{cpWER}, where all transcriptions of one speaker are concatenated and the word error rate is computed on the mapping between estimated transcriptions and the ground-truth word sequences of the different speakers that results in the lowest word error rate.

Note that the \gls{cpWER} also counts recognition errors caused by wrong diarization.
For example, if the diarization swaps the speaker identity of an utterance, this adds to the deletion error count of the correct speaker's transcription, and, furthermore, adds to the insertion error count of the wrongly identified speaker, regardless of whether the recognized words were correct or not.

\glsreset{DIcpWER}
Since we are interested in knowing %
the contribution of diarization errors, we also computed  %
the assignment $k_b$ of recognized segment transcriptions to the ground-truth transcription streams such that the \gls{cpWER} is as small as possible. In other words, we use as the speaker label for each segment $b$ the label that minimizes the \gls{cpWER}, instead of the estimated speaker label $k_b$ that was obtained from $\diaEst$, while keeping the estimated segment start and end times.
This is similar to the MIMO-WER defined in \cite{vonNeumann2022WER}, but ground-truth and estimate are swapped.
The error rate computed in this way ignores contributions from diarization errors, hence we call it \gls{DIcpWER}.
The difference between the \gls{cpWER} and \gls{DIcpWER} can then be interpreted as the errors caused by diarization mistakes.

\glsreset{CSS}

In the literature, the asclite tool is often used to compute the \gls{WER} of multi-speaker transcriptions \cite{Fiscus2006asclite}.
It ignores diarization errors as well, but is limited to shorter segments and fewer estimated \enquote{speakers}.  It has been used on LibriCSS in the past, however with so-called \gls{CSS} systems that separate the data into two streams 
where each stream contains the signals of several speakers who do not overlap %
\cite{Chen2020LibriCSS}.
In our setup, we produce eight streams, one for each speaker in the meeting, and because the computation and memory demands of asclite increase exponentially with the number of streams, it  is computationally infeasible to use asclite for our system.

Both asclite and \gls{DIcpWER} calculate a \gls{WER} where diarization errors are ignored. The numbers obtained are thus roughly comparable, and we shall use \gls{SAgWER} as a general name for those \glspl{WER} that ignore diarization errors.

We used two different pretrained \gls{ASR} systems from the ESPnet framework \cite{Watanabe2018_ESPnetEndtoEndSpeech} to estimate the transcription.
Both are used ``out of the box'' without any fine-tuning.
The first, which we refer to as ``base''\footnote{\url{https://zenodo.org/record/3966501}} \cite{watanabe2020PretrainedASR}, has a transformer architecture.
It is designed and trained on LibriSpeech and achieves a \gls{WER} of \SI{2.7}{\percent} on the clean test set of LibriSpeech.
The second, which we refer to as ``WavLM''\footnote{\url{https://huggingface.co/espnet/simpleoier_librispeech_asr_train_asr_conformer7_wavlm_large_raw_en_bpe5000_sp}} \cite{Chang2022WavLMASR},
is a conformer based \gls{ASR} system that utilizes WavLM \cite{Chen2022wavlm} features.
It is designed for CHiME-4 \cite{Vincent2017CHiME4} and uses Libri-Light \cite{Kahn2020librilight}, GigaSpeech \cite{Chen2021gigaspeech}, VoxPopuli \cite{Wang2021voxpopuli}, CHiME-4 \cite{Vincent2017CHiME4}, and WSJ0/1 \cite{Paul1992WSJ} for training.
On the clean test set of LibriSpeech, it achieves a \gls{WER} of \SI{1.9}{\percent}.

The results in the following tables have been obtained using all seven channels of the LibriCSS dataset, unless otherwise stated.
Note that masks are estimated for each channel independently, and the median is then applied across the channels to obtain the final mask, see \cref{fig:tssep}. %

\subsection{Preprocessing}
The speech data is transformed to the \gls{STFT} domain with a \SI{64}{ms} window and \SI{16}{ms} shift. In the \gls{STFT} domain, the logarithmic spectrogram and \gls{MFCC} are stacked as input features for the \gls{NN}.

For the computation of the speaker embeddings, we followed \cite{Medennikov20TS-VAD}:
at training time, we used oracle annotations to identify single-speaker regions, from which I-vectors are estimated;
at test time, we employed an X-vector extractor and \gls{SC} from the ESPnet LibriCSS recipe \cite{Watanabe2018_ESPnetEndtoEndSpeech} to obtain estimates of the annotations, to be used for I-vector extraction\footnote{\url{https://kaldi-asr.org/models/13/0013_librispeech_v1_extractor.tar.gz}}.
Note that this initial estimate is unable to identify overlapping speech, and thus overlapping speech cannot be excluded for the embedding calculation.

In Kaldi \cite{Povey2011Kaldi}, and hence also in the original implementation of \gls{TSVAD} \cite{Medennikov20TS-VAD}, it is common to compute a mean and variance vector for each data set and normalize the input feature vectors with them.
We initially skipped this normalization, but we observed a serious performance degradation without this domain adaptation, with an increase of the \gls{WER} 
from \SI{7.7}{\percent} to \SI{21.5}{\percent}.
Therefore, for all experiments reported here, we employed moment (mean and variance) matching between the simulated validation dataset and the rerecorded DEV and EVAL LibriCSS datasets for the input features and the I-vectors.

As an additional enhancement, \gls{WPE} based dereverberation \cite{Nakatani2010WPE, Drude2018nara} was carried out prior to the \gls{NN} and the extraction at test time.

\subsection{Training Details}
\label{sec:trng_details}

While developing this system, we found several aspects that had a higher impact on the convergence properties or the final performance than we expected, for example reducing the time until the network starts to learn from weeks to hours.
We will not provide detailed experiments for each of the modifications we found useful, but we want to  present our findings for the benefit of the community.

We used one-minute-long chunks of data for the training.
To reduce the dependency on the artificial speaker ordering, we followed \cite{Medennikov20TS-VAD} in randomly permuting the speaker embeddings at the input and reversing the permutation after the \gls{NN}.
Furthermore, we observed a better training convergence when executing the \gls{NN} two times with different permutations and averaging the outputs, after inverting the permutations, before the last nonlinearity.
This idea was mentioned in \cite{Medennikov20TS-VAD} to be used for the evaluation of some preliminary experiments.

\label{sec:mixup}
For separation systems, we observed better convergence properties when more overlap was present in the beginning of the training.
We also observed this behavior for \gls{TSVAD} based diarization, hence we used the superposition idea from \cite{Ebbers2021DCASE}, which is similar to mixup \cite{Zhang2017Mixup} and took, with $\zeta = \SI{50}{\percent}$ probability, the sum of a minute of data with the following minute as training data instead of the original data.
To compute  the \gls{TSVAD} targets for this superposition, we used the logical {\tt or} operator.
Although this superposition introduced a mismatch between training and test (e.g., self overlap and more simultaneous and total active speakers), we observed better convergence and a better final performance.
We show the distribution of the number of active speakers in the training data with and without superposition in \cref{fig:hist:speakers}.

\subsection{Postprocessing / Segmentation}
\label{sec:postprocessing}

For the pretrained \gls{ASR} system named  ``base'', we observed a strange behavior for long segments\footnote{e.g., a transcription suddenly interrupted by a stream of single-letter words consisting of the letters E, O, and T.}, hence we split long activities at silence positions, according to $\diaProb$, such that no segment is longer than \SI{12}{s}.
As minimum segment length, we used 40 frames (\SI{0.64}{s}), but this was only necessary when no overestimation was used.

\begin{table}[t]
    \centering
    \caption{Comparison of our TS-VAD implementation with that of \cite{He21TS-VAD}, using various diarization information to segment the observation and applying various \gls{ASR} systems, without intermediate separation of overlapping speech.
    The first microphone of LibriCSS is used.
    } 
    \label{tab:res:dia}
    \begin{tabular}{l@{~~}l@{~~}l l S@{~~}S S@{~~}S}
        \toprule
         & & & & \multicolumn{2}{c}{\textbf{\acrshort{DER}}} & \multicolumn{2}{c}{\textbf{\acrshort{cpWER}}} \\
         \cmidrule(lr){5-6} \cmidrule(lr){7-8}
         \bf ID & \bf Ref. & \bf Diarization & \bf ASR & {DEV} & {EVAL} & {DEV} & {EVAL} \\
         \midrule
         (1) & \cite{He21TS-VAD} & Oracle  & HMM-DNN & {--} & {--} & {--} & 23.1 \\
         (2) & \cite{He21TS-VAD} & TS-VAD  & HMM-DNN & {--} & 7.6 & {--} & 25.8 \\
         (3) & \cite{Chen2022wavlm} & Oracle  & WavLM & {--} & {--} & {--} & 6.0 \\
         \midrule
         (4) & & Oracle  & Base & 0.0 &       0.0 &      19.17 &       19.77 \\
         (5) & & SC & Base & 18.93 & 17.93 & 29.07 & 27.77  \\
         (6) & & TS-VAD & Base & 6.30 &      5.65 &      21.55 &       21.27 \\  %
         (7) & & TS-VAD & WavLM & 6.30 &      5.65 &      10.39 &        9.13 \\ %
         (8) & & Oracle & WavLM & 0.0 &   0.0 &    6.79 &        6.01 \\
         \bottomrule
    \end{tabular}
    \label{tab:my_label}
\end{table}

\subsection{Baselines and reference results}

The public source code of TS-VAD \cite{Medennikov20TS-VAD} is written in Kaldi for the CHiME-6 data, and the modifications for LibriCSS \cite{He21TS-VAD} are not publicly available.
We reimplemented \gls{TSVAD} in PyTorch \cite{Paszke2019PyTorch} and tried to follow the original implementation as closely as possible. For example, for the \gls{NN} layers, we used one BLSTMP\footnote{A BLSTMP is a bidirectional long short-term memory \cite{Hochreiter1997LSTM} recurrent neural network with a feedforward projection layer.} \cite{Sak2015LSTMP}, two BLSTMP, and one BLSTMP followed by a feedforward layer for the three \gls{NN} blocks in \cref{fig1}.
Nevertheless, they may differ in some details. We therefore first verify if our implementation is competitive.
In \cref{tab:res:dia}, we see that the \gls{DER} and \gls{WER} of our implementation is similar to the original implementation (lines (6) vs.\ (2)).
The \gls{TSVAD} is able to improve  over \gls{SC} (lines (6) vs.\ (5)) and, with the WavLM-based \gls{ASR} system, we achieved our best \gls{WER} without doing any enhancement (line (7)).

\begin{table}[t]
\centering

  \caption{
    Results of TS-SEP with various enhancement strategies, using segmentation hyperparameters $\text{Thr.} \tau = 0.6$, $\text{Dil.} = 161$, and $\text{Ero.}=81$.
  }
  \label{tab:res:sep:enh}
  \begin{tabular}{c@{~}H c c@{~}c S@{~~~}S S@{~~~}S }
    \toprule
         & & & \multicolumn{2}{c}{\textbf{Extraction}} & \multicolumn{2}{c}{\textbf{\acrshort{DER}}} & \multicolumn{2}{c}{\textbf{\acrshort{cpWER}}} \\
        \cmidrule(lr){4-5} \cmidrule(lr){6-7} \cmidrule(lr){8-9}
        \bf ID & \textbf{Init} & \textbf{WPE} & ~{BF}~ & {Masking ($\thrMasking$)} & {DEV} & {EVAL} & {DEV} & {EVAL}  \\
        \midrule
        (1) & SC & \xmark & \xmark & \xmark & 15.47 &     13.89 &   26.53 &    24.42 \\  %
        (2) & SC & \cmark & \xmark & \xmark & 15.35 &     13.68 &   26.26 &    24.08 \\  %
        (3) & SC & \cmark & \xmark & \cmark (0.0) & 15.35 &     13.68 &   12.42 &    10.49 \\  %
        (4) & SC & \cmark & \xmark & \cmark (0.5) & 15.35 &     13.68 &   22.99 &    21.07 \\  %
        (5) & SC & \cmark & \cmark & \xmark & 15.35 & 13.68 & 10.51 & 8.82	 \\  %
        \midrule
        (6) & SC & \cmark & \cmark & \cmark (0.0) & 15.35 & 13.68 & 11.08 & 9.29 \\  %
        (7) & SC & \cmark & \cmark & \cmark \rlap{(0.01)}\hphantom{(0.0)} & 15.35 & 13.68 & 11.12 & 9.26 \\  %
        (8) & SC & \cmark & \cmark & \cmark \rlap{(0.05)}\hphantom{(0.0)} & 15.35 & 13.68 & 10.93 & 9.00	 \\  %
        (9) & SC & \cmark & \cmark & \cmark (0.2) & 15.35 & 13.68 & 10.43 & 8.48	 \\  %
        (10) & SC & \cmark & \cmark & \cmark (0.5) & 15.35 & 13.68 & \bf 10.16 & \bf 8.42 \\  %
        \midrule
        (11) & SC & \xmark & \cmark & \cmark (0.5) & 15.47 &     13.89 &   11.31 &     9.46 \\  %
    \bottomrule
  \end{tabular}
\end{table}

\subsection{Extraction and enhancement}
In \cref{tab:res:sep:enh}, we analyze the performance of our proposed TS-SEP, which computes a mask to extract the source signals of the speakers from the observation, under various extraction strategies.
We first use no enhancement (line (1)) to see the diarization performance.
Note that we used a dilation of $161$ frames in \cref{tab:res:sep:enh} instead of the $81$ frames used in \cref{tab:res:dia}.
The higher \gls{DER} is caused by the resulting overestimation, and the \gls{cpWER} gets slightly worse compared to \gls{TSVAD} (we explain why in the next section).
Including \gls{WPE}-based dereverberation slightly improved the performance (line (2)), and using masking for enhancement gave a big jump down to \SI{10.49}{\percent} \gls{cpWER} (line (3)).
With beamforming (line (5)), we were able to further improve the \gls{cpWER} to \SI{8.82}{\percent}.
Doing masking after beamforming (line (6)) yielded worse results, probably due to the artifacts introduced by masking.
But by selecting a proper lower threshold $\thrMasking=0.5$ for the mask, we were able to obtain a slight improvement and achieved \SI{8.42}{\percent} \gls{cpWER} (line (10)).
\diff{}{We also include line (11) to help illustrate the effect of WPE.}

\diff{}{The effect of the lower threshold is in line with the observations made in\mbox{\cite{iwamoto22_interspeech}} for speech enhancement}\footnote{\diff{}{In\mbox{\cite{iwamoto22_interspeech}}, a scaled version of the observation is added to the estimate, which in our system corresponds to adding a constant to the mask. Note that we do not add the observation, since it contains the cross-talker and beamformed signals are known to work well for \gls{ASR}\mbox{\cite{Heymann_Drude_Haeb-Umbach_2016}}.}}\diff{}{: for \gls{ASR}, the artifacts produced by masking have a negative effect, but using a carefully chosen lower threshold for the mask improves the \gls{ASR} performance.}

\subsection{Segmentation}
\label{sec:ex:seg}

In preliminary experiments not reported here, we optimized the segmentation parameters to minimize the \gls{DER} when using a \gls{TSVAD} system.
With the \gls{TSSEP} system, two things changed.
First,  an overestimation
of the interval length of activity might be non-critical for source extraction, because potentially active cross-talkers at the borders can be suppressed by the enhancement.
Further, an overestimation
yields utterance boundaries that better match the typical training data of  \gls{ASR} systems, as mentioned in \cref{sec:segmentatuon}.
Second, \gls{TSSEP} is trained to predict a reconstruction mask, so the activity estimates $\diaProb$ are often smaller than with \gls{TSVAD}, especially in TF bins where the speaker is inactive or less active, as shown in \cref{fig:activity}.
The \gls{TSVAD} system, on the other hand, is trained to yield values close to one for the full duration of an utterance.
Because of this training mismatch, we expected \gls{TSSEP} to prefer a smaller threshold $\thrDia$ in \cref{eq:closing}.%

\begin{table}[t]
\centering
  \caption{
    Results of TS-SEP with various segmentation hyperparameters, using WPE+BF+Masking~($\thrMasking=0.5$) as enhancement.
  }
  \label{tab:res:sep:seg}
  \begin{tabular}{c c@{~}c@{~}c H H S@{~~~}S S@{~~~}S }
    \toprule
         & \multicolumn{3}{c}{\textbf{Segmentation}} & & & \multicolumn{2}{c}{\textbf{\acrshort{DER}}} & \multicolumn{2}{c}{\textbf{\acrshort{cpWER}}} \\
        \cmidrule(lr){2-4} \cmidrule(lr){5-8} \cmidrule(lr){9-10}
        \bf ID & Thr. $\thrDia$ & Dil. & Ero. & BF & Mask. & {DEV} & {EVAL} & {DEV} & {EVAL} \\
    \midrule
    (1) & 0.3 & 201& 81  & \cmark & \cmark & 24.91 & 21.94 & 10.41 &  8.26 \\  %
    (2) & 0.3 & 161 & 81 & \cmark & \cmark & 17.23 & 14.69 & 10.06 &  7.72 \\  %
    (3) & 0.3 & 121 & 81 & \cmark & \cmark &  9.89 &  7.89 &  9.99 &  7.70 \\  %
    (4) & 0.3 & 81 & 81  & \cmark & \cmark &  7.61 &  6.49 & 12.98 & 11.16 \\  %
    \midrule
    (5) & 0.1 & 161 & 81 & \cmark & \cmark & 19.55 &    16.46 &   10.36 &     7.98 \\  %
    (6) & 0.2 & 161 & 81 & \cmark & \cmark & 18.22 &    15.38 &   10.27 &     7.79 \\  %
    (8) & 0.3 & 161 & 81 & \cmark & \cmark & 17.23 &     14.69 &   10.06 &     7.72 \\  %
    (9) & 0.4 & 161 & 81 & \cmark & \cmark & 16.44 &    14.16 &  \bf  9.95 &  \bf   7.72 \\  %
    (10) & 0.6 & 161 & 81 & \cmark & \cmark & \bf 15.35 &   \bf 13.68 &   10.16 &     8.42 \\  %
    \midrule
    (11) & 0.3 & 141 & 61 & \cmark & \cmark & \bf 17.14 &  \bf  14.61 & \bf  10.01 &  \bf   7.72 \\  %
    (12) & 0.3 & 161 & 81 & \cmark & \cmark & 17.23 &     14.69 &   10.06 &     7.72 \\  %
    (13) & 0.3 & 181 & 101 & \cmark & \cmark & 17.41 &    14.85 &   10.06 &     7.75 \\  %
    \bottomrule
  \end{tabular}
\end{table}

We verified both hypotheses in the experiments of \cref{tab:res:sep:seg}.
When optimizing the segmentation hyperparameters, the \gls{DER} and \gls{WER} showed opposite behavior.
For the \gls{DER}, as expected, an overestimation has a negative effect (lines (4) vs.\ (1) to (3)).
For the \gls{WER}, on the contrary, some overestimation is helpful, as argued in the previous paragraph.
While in preliminary experiments a threshold of $\thrDia = 0.6$ was a good choice for \gls{TSVAD}, smaller values in the range of $0.3$ to $0.4$ led to best performance for \gls{TSSEP} (lines (5) to (10)).
Simultaneously expanding or reducing the dilation and erosion windows had only a minor effect (lines (11) to (13)).

\begin{table*}[th]
\centering
  \newcommand{\singleCH}{\cmark\xspace}
  \newcommand{\multiCH}{\xmark\xspace}
  \caption{
    Towards a single-channel system.
    A \singleCH indicates that a component is single-channel and a \multiCH indicates that the component uses multiple channels. DA means domain adaptation, where we distinguish whether it is done for all channels simultaneously (\multiCH) or per channel (\singleCH).
  }
   \label{tab:res:sep:side}
  \begin{tabular}{llllSSSSSSSS}
    \toprule
    & & & & \multicolumn{4}{c}{\textbf{\acrshort{cpWER}}} & \multicolumn{4}{c}{\textbf{\acrshort{DIcpWER}}} \\
    \cmidrule(lr){5-8} \cmidrule(lr){9-12}
    {} & & & & \multicolumn{2}{c}{{{DEV}}} & \multicolumn{2}{c}{{{EVAL}}} & \multicolumn{2}{c}{{{DEV}}} & \multicolumn{2}{c}{{{EVAL}}} \\
    \cmidrule(lr){5-6} \cmidrule(lr){7-8} \cmidrule(lr){9-10} \cmidrule(lr){11-12}
    \bf ID & \bf WPE & \smash{\tab[c][b]{\bf Mask \\\bf Reduction}} &  \bf Extraction & {DA: \multiCH} & {DA: \singleCH} & {DA: \multiCH} & {DA: \singleCH} & {DA: \multiCH} & {DA: \singleCH} & {DA: \multiCH} & {DA: \singleCH} \\
    \midrule
    (1) &   \multiCH &      \multiCH~ (median) &  \multiCH~ (BF+Masking ($\thrMasking=0.5$)) &   10.06 &        9.83 &     7.72 &         7.70 &         5.41 &             5.12 &          5.92 &              5.85 \\
    (2) &   \multiCH &  \singleCH~(reference) &  \multiCH~ (BF+Masking ($\thrMasking=0.5$)) &    8.96 &        8.45 &     9.05 &         7.80 &         5.94 &             5.37 &          6.59 &              5.98 \\
    (3) &   \multiCH &      \multiCH~ (median) &    \singleCH~(Masking ($\thrMasking=0.0$)) &   12.14 &       11.90 &    10.11 &        10.00 &         7.55 &             7.27 &          8.34 &              8.20 \\
    (4) &   \multiCH &  \singleCH~(reference) &    \singleCH~(Masking ($\thrMasking=0.0$)) &   11.83 &       11.07 &    12.04 &        10.30 &         8.80 &             8.07 &          9.70 &              8.51 \\
    (5) &  \singleCH &  \singleCH~(reference) &    \singleCH~(Masking ($\thrMasking=0.0$)) &   13.40 &       12.12 &    13.27 &        11.55 &        10.45 &             9.01 &         10.83 &              9.74 \\
    \bottomrule
  \end{tabular}
  
\end{table*}

\subsection{Single-channel system}
In \cref{tab:res:sep:side}, we replaced components of our system with single-channel counterparts to
end up with a single-channel system: beamforming, masking, dereverberation, and domain adaptation (DA).
Using beamforming obviously requires multiple channels, so the single-channel alternative is simply not to use any beamforming.
By contrast, a mask for source extraction can be computed either from a single (reference) channel or in a multi-channel fashion by applying the mask estimation \gls{NN} independently to each channel and reducing the resulting masks to a single mask with the median operation.
Similarly, \gls{WPE}-based dereverberation is usually done jointly over all available channels, but can also be applied to each channel independently.
For the domain adaptation, we used by default one channel of the simulated validation set to compute the training time statistics, and all channels from the rerecorded test set to compute the test time statistics.
To make it single-channel, we compute the test time statistics independently on each channel.
For the DEV set, we observed some outliers for the \gls{cpWER}\footnote{The DEV set was chosen for diarization-agnostic metrics, where 6 recordings are enough. On diarization-aware metrics, one wrong assignment can have a large impact because of the set's small size.}, hence we report here also the \gls{DIcpWER}, which is more robust.

Switching one component after each other to the single-channel variant increased the \gls{cpWER}, except for the domain adaptation, which had a positive effect. 
We investigated this surprising result with more experiments.
It turned out that on the LibriCSS data, which are true recordings with a circular microphone array including a microphone in the center of the circle,  the reference microphone, which was the center microphone, had notably different statistics compared to the other microphones. 
This is different from the simulated data, which  employed omnidirectional microphone characteristics, where no relevant differences in the statistics between the microphones were observable.
With a completely single-channel system, we finally arrive at a \gls{cpWER} of \SI{11.55}{\percent} and a \gls{DIcpWER} of \SI{9.74}{\percent}.

\begin{table*}[th]
\newcommand*\circled[1]{\smash{\tikz[baseline=(char.base)]{
		\node[rounded rectangle, fill=black,text=white,inner xsep=0.0ex,inner ysep=0.2ex] (char) {#1};}}}
\centering
  \caption{
    \gls{TSVAD} vs \gls{TSSEP} with different extractions and ASR systems, \\
    where $\text{Dil.}=161$, $\text{Ero.}=81$ and $\text{Thr.}$ is set to $\tau=0.6$ (\gls{TSVAD}) or $\tau=0.3$ (\gls{TSSEP}). 
  }
  \label{tab:res:sep:gss}
  \begin{tabular}{c l l l  S@{~~~}S  S@{~~~}S  S@{~~~}S }
    \toprule
     & &  & & \multicolumn{2}{c}{\bf \gls{DER}} & \multicolumn{2}{c}{\bf \gls{cpWER}} & \multicolumn{2}{c}{\bf \gls{DIcpWER}} \\
     \cmidrule(lr){5-6}
     \cmidrule(lr){7-8}
      \cmidrule(lr){9-10}
    \bf ID & \bf Dia./Sep. & \bf Extraction & \bf ASR & {DEV} & {EVAL} & {DEV} & {EVAL} & {DEV} & {EVAL} \\
    \midrule
    (1) & TS-VAD & BF+Masking ($\thrMasking=0.5$)   &   Base &    18.72 &     16.28 &      17.26 &       15.98 &         13.22 &          13.87 \\
    (2) & TS-VAD & GSS T-init &   Base &    18.72 &     16.28 &       9.09 &        6.43 &          4.74 &           4.32 \\
    \midrule
    (3) & TS-SEP & BF+Masking ($\thrMasking=0.5$)        &   Base &    17.23 & 14.69 &      10.06 &        7.72 &          5.41 &           5.92 \\
    (4) & TS-SEP & GSS T-init &   Base &    17.23 &     14.69 &       8.93 &        6.32 &          4.23 &           4.54 \\
    (5) & TS-SEP & GSS TF-init &   Base &    17.23 &     14.69 &       8.80 &        6.15 &          4.07 &           4.37 \\
    \midrule
    (6) & TS-VAD & BF+Masking ($\thrMasking=0.5$)   & WavLM &    18.72 &     16.28 &      12.13 &        9.50 &          7.95 &           7.35 \\
    (7) & TS-VAD & GSS T-init & WavLM &    18.72 &     16.28 &       8.44 &        5.47 &          4.20 &           3.32 \\
    \midrule
    (8) & TS-SEP & BF+Masking ($\thrMasking=0.5$)       & WavLM &    17.23 &     14.69 &       8.55 &        5.76 &          3.85 &           3.92 \\
    (9) & TS-SEP & GSS T-init      & WavLM &    17.23 &     14.69 &       8.12 &        5.10 &          3.40 &           3.29 \\
    (10) & TS-SEP & GSS TF-init & WavLM &    17.23 &     14.69 &  \bf     8.11 &    \bf    5.06 &    \bf      3.32 &           \bf 3.26 \\
    \bottomrule
  \end{tabular}
  
\end{table*}
  
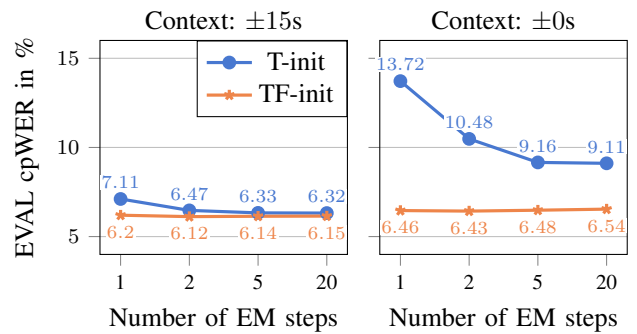
\begin{figure}
     \centering
\begin{tikzpicture}

\definecolor{darkslategray38}{RGB}{38,38,38}
\definecolor{darkslategray66}{RGB}{66,66,66}
\definecolor{lightgray204}{RGB}{204,204,204}
\definecolor{peru21713894}{RGB}{217,138,94}
\definecolor{steelblue89124191}{RGB}{89,124,191}

\definecolor{royalblue72120208}{RGB}{72,120,208}
\definecolor{coral23813374}{RGB}{238,133,74}

\pgfplotstableread[row sep=\\,col sep=&]{
 UseMask & iterations &  context &  WER DEV &  WER eval &  MIMO-WER dev &  MIMO-WER eval \\
   False &          1 &   240000 &     9.88 &      7.11 &          5.16 &           5.30 \\
   False &          2 &   240000 &     9.12 &      6.47 &          4.53 &           4.71 \\
   False &          5 &   240000 &     8.95 &      6.33 &          4.29 &           4.55 \\
   False &         20 &   240000 &     8.93 &      6.32 &          4.23 &           4.54 \\
    }\mydata
    
\pgfplotstableread[row sep=\\,col sep=&]{
 UseMask & iterations &  context &  WER DEV &  WER eval &  MIMO-WER dev &  MIMO-WER eval \\
    True &          1 &   240000 &     8.95 &      6.20 &          4.23 &           4.41 \\
    True &          2 &   240000 &     8.93 &      6.12 &          4.25 &           4.34 \\
    True &          5 &   240000 &     8.84 &      6.14 &          4.18 &           4.35 \\
    True &         20 &   240000 &     8.80 &      6.15 &          4.07 &           4.37 \\
    }\mydataD
    
\pgfplotstableread[row sep=\\,col sep=&]{
 UseMask & iterations &  context &  WER DEV &  WER eval &  MIMO-WER dev &  MIMO-WER eval \\
   False &          1 &        0 &    16.77 &     13.72 &         12.20 &          11.87 \\
   False &          2 &        0 &    13.72 &     10.48 &          8.90 &           8.64 \\
   False &          5 &        0 &    12.33 &      9.16 &          7.49 &           7.32 \\
   False &         20 &        0 &    12.06 &      9.11 &          7.30 &           7.29 \\
    }\mydataDD
    
\pgfplotstableread[row sep=\\,col sep=&]{
 UseMask & iterations &  context &  WER DEV &  WER eval &  MIMO-WER dev &  MIMO-WER eval \\
    True &          1 &        0 &     9.43 &      6.46 &          4.67 &           4.66 \\
    True &          2 &        0 &     9.25 &      6.43 &          4.47 &           4.64 \\
    True &          5 &        0 &     9.35 &      6.48 &          4.61 &           4.67 \\
    True &         20 &        0 &     9.37 &      6.54 &          4.60 &           4.73 \\
    }\mydataDDD

\begin{axis}[
            width=0.55\columnwidth,
            height=.5\columnwidth,
            legend style={at={(1,1)},
                anchor=north east
                },
            xtick=data,
            ymajorgrids,
            tick align=outside,
            tick pos=left,
            symbolic x coords={1,2,5,20},
            nodes near coords,
            nodes near coords align={vertical},
            nodes near coords style={fill=white, inner sep=0.1ex, outer sep=0.8ex, font=\scriptsize},
            ymin=4,ymax=16,
            ylabel={EVAL cpWER in \%},
            xlabel={Number of EM steps},
            ticklabel style = {font=\footnotesize},
            title={Context: $\pm 15 \si{s}$},
            every axis title/.style={at={(0.5,1)},anchor=south,outer sep=0.8ex,inner sep=0ex},
        ]

        \addplot[royalblue72120208, very thick, mark=*] table[x=iterations,y=WER eval]{\mydata};
        \addplot[coral23813374, very thick, mark=star, every node near coord/.append style={anchor=north}] table[x=iterations, y=WER eval]{\mydataD};

        \legend{T-init, TF-init}
    \end{axis}

\pgfplotstableread[row sep=\\,col sep=&]{
 UseMask & iterations &  context &  WER DEV &  WER eval &  MIMO-WER dev &  MIMO-WER eval \\
   False &          1 &        0 &    16.77 &     13.72 &         12.20 &          11.87 \\
   False &          2 &        0 &    13.72 &     10.48 &          8.90 &           8.64 \\
   False &          5 &        0 &    12.33 &      9.16 &          7.49 &           7.32 \\
   False &         20 &        0 &    12.06 &      9.11 &          7.30 &           7.29 \\
    }\mydataDD
    
\pgfplotstableread[row sep=\\,col sep=&]{
 UseMask & iterations &  context &  WER DEV &  WER eval &  MIMO-WER dev &  MIMO-WER eval \\
    True &          1 &        0 &     9.43 &      6.46 &          4.67 &           4.66 \\
    True &          2 &        0 &     9.25 &      6.43 &          4.47 &           4.64 \\
    True &          5 &        0 &     9.35 &      6.48 &          4.61 &           4.67 \\
    True &         20 &        0 &     9.37 &      6.54 &          4.60 &           4.73 \\
    }\mydataDDD

\begin{axis}[
            xshift=0.42\columnwidth,
            width=0.55\columnwidth,
            height=.5\columnwidth,
            legend style={at={(1,1)},
                anchor=north east
                },
            xtick=data,
            ymajorgrids,
            tick align=outside,
            tick pos=left,
            symbolic x coords={1,2,5,20},
            nodes near coords,
            nodes near coords align={vertical},
            nodes near coords style={fill=white, inner sep=0.1ex, outer sep=0.8ex, font=\scriptsize},
            ymin=4,ymax=16,
            yticklabels={,,},
            xlabel={Number of EM steps},
            ticklabel style = {font=\footnotesize},
            title={Context: $\pm 0 \si{s}$},
            every axis title/.style={at={(0.5,1)},anchor=south,outer sep=0.8ex,inner sep=0ex},
        ]

        \addplot[royalblue72120208, very thick, mark=*] table[x=iterations,y=WER eval]{\mydataDD};
        \addplot[coral23813374, very thick, mark=star, every node near coord/.append style={anchor=north}] table[x=iterations,y=WER eval]{\mydataDDD};

    \end{axis}
\end{tikzpicture}%
     \caption{\gls{TSSEP} with \gls{GSS} for different numbers of guided \gls{EM} steps and different amounts of temporal context, where \gls{VAD} (T-init) or mask initialization (TF-init) is used for \gls{GSS}.}
     \label{fig:gss:contextSteps}
 \end{figure}

\subsection{TS-VAD vs.\ TS-SEP and mask refinement with GSS}

\Gls{GSS} can be considered to be an alternative to \gls{TSSEP} to expand voice activity information from time to time-frequency resolution.
However, it can also be used as a mask refinement operation to be applied to the \gls{TSSEP} output.
In \cref{tab:res:sep:gss}, we explore these two different setups.

To start with, we compared the performance of \gls{TSVAD} with \gls{TSSEP} without postprocessing their outputs with \gls{GSS} (lines (1) and (3)).
We see a clear advantage of using a mask with time-frequency resolution for the \gls{cpWER} (\SI{7.72}{\%}  vs.\ \SI{15.98}{\%}).
Next, only temporal activity information (T-init) instead of a time-frequency mask are used from \gls{TSSEP} (same for \gls{TSVAD}, naturally).
Expanded by \gls{GSS} to a time-frequency mask, the mask is used for source extraction (lines (2) and (4)).
Both configurations benefit from \gls{GSS}, however \gls{TSVAD} benefits considerably more than \gls{TSSEP}, such that the margin between them is drastically reduced to \SI{6.43}{\%} vs.{} \SI{6.32}{\%} \gls{cpWER}.
In contrast to \gls{TSVAD}, \gls{TSSEP} can utilize \gls{GSS} as mask refinement (TF-init), which slightly improves the \gls{cpWER} to \SI{6.15}{\%}.

For \gls{GSS}, we  used a context length of $\pm$\SI{15}{s}, such that each segment to which the EM algorithm is applied is expanded by \SI{30}{s}.
While it appears that the advantage of \gls{TSSEP} over \gls{TSVAD} is almost lost if \gls{GSS} is used as a postprocessing step, this conclusion should be reconsidered in light of the high computational complexity of \gls{GSS}\footnote{For example in the CHiME-6 baseline \cite{Watanabe2020CHiME6}, \gls{GSS} used 12 microphones instead of 24 microphones, because the runtime was too large.}. 
In \cref{fig:gss:contextSteps}, we considered different initialization based on the output of \gls{TSSEP} and different context lengths, and  used up to 20 guided and one non-guided \gls{EM} steps, which are the default.
It can be seen that mask-based initialization (TF-init) of \gls{GSS} is always better than temporal activity based initialization (T-init).
Further, using the mask allows one to reduce the number of \gls{EM} steps without a relevant effect on the \gls{WER}.
That fewer \gls{EM} steps are necessary is expected, since the mask is already a solid starting point. 

Further, the context by which the segments are expanded can be significantly reduced with mask initialization.
This can also be well explained by the ambiguity incurred if two speakers overlap for the whole duration of a segment:
while a mask-based TF-level initialization can then still distinguish between the speakers, the temporal activity information is sometimes insufficient to do so, and left and right contexts need to be added to have a greater chance that 
the activity patterns between speakers are sufficiently different.
See \cite{Boeddeker2018GSS} for a detailed discussion on the virtue of context expansion for \gls{GSS}.

\Cref{tab:res:sep:gss} further contains  results with the WavLM model, from which similar conclusions can be drawn. Using the stronger WavLM-based \gls{ASR} system, we obtained our best \gls{cpWER} of \num{5.06}\% for \gls{TSSEP}.

Note that the table also includes the \gls{DIcpWER} results, which is speaker agnostic and thus does not count the contribution of diarization errors to the \gls{WER}.
Here, the best configuration achieves a WER of \num{3.26}\%.

\begin{table}
\centering
    \makeatletter \DeclareRobustCommand{\arrow}{%
\check@mathfonts%
\raisebox{\fontdimen22\textfont2}{\tikz{\draw[arrow](0,0)--(0.8em,0);}}%
\xspace}\makeatother

    \caption{
        Literature comparison. The style column indicates how the system is designed, where D, S, and A signify diarization, separation, and ASR, respectively. An \arrow indicates a sequence and a $+$ indicates a coupling between components. The $\dagger$ indicates that the \gls{cpWER} is calculated on roughly $1$~minute chunks of the $10$~minute data.
    }
      \label{tab:res:final}

  \begin{tabular}{@{~}l@{~~}l@{~}c@{~}S@{}l@{~}S@{~}}
    \toprule
    \bf \tab{\\System (other)}  & \bf \tab{\\Style} & \bf  \tab{Single\\Ch.} & \bf  \tab{\\cpWER}%
    && \bf  \tab{SAg-WER\\asclite} \\
    \midrule
    CSS with DOA Dia \cite{Wang2022DOACSS} & S \arrow D \arrow A & \xmark & 12.98 & & {\xmark} \\
    CSS with DOA Dia \cite{Wang2022DOACSS} & S \arrow D \arrow A & \xmark & 12.40 & $\dagger$ & {\xmark} \\
    SMM \cite{boeddeker22_interspeech} & D \arrow S \arrow A & \xmark &  5.9 & $\dagger$ & {\xmark} \\
    \midrule
    CSS \arrow SC \cite{Raj2021Meeting} & S \arrow D \arrow A & \xmark & 12.7 & & {\xmark} \\
    \midrule
    t-SOT TT \cite{Kanda22tSOT} & E2E & \cmark & {\xmark} & & 7.6 \\
    CSS \cite{Wang2021css} & S \arrow A & \cmark & {\xmark} & & 10.15 \\ %
    CSS \cite{Wang2021css} & S \arrow A & \xmark & {\xmark} & & 5.85 \\ %
    \midrule
    Transcribe-to-Diarize \cite{Kanda2022T2D} & E2E & \cmark &  11.6 & & {\xmark} \\
    TS-VAD + Speakerbeam \cite{Delcroix2021MeetinSpeakerBeam} & D \arrow S \arrow A & \cmark & 18.8 & & {\xmark} \\
    TS-VAD + GSS \cite{Delcroix2021MeetinSpeakerBeam} & D \arrow S \arrow A & \xmark & 11.2 & & {\xmark} \\
    SC + GSS \cite{Raj2022gpuGSS} & D \arrow S \arrow A & \xmark & 12.12 & & {\xmark} \\
    \midrule
    \tblrowadded\bf \tab{\\System (mixed)} &  \bf \tab{\\Style} & \tab{\bf Single\\\bf Ch.} & \bf \tab{\\cpWER} %
    && \tab{\bf SAg-WER\\\bf DI-cpWER} \\
    \midrule
    \tblrowadded SC \cite{Raj2022gpuGSS} \arrow GSS \arrow WavLM & D \arrow S \arrow A & {\xmark} & 13.85 && 8.82 \\
    \midrule
    \bf \tab{\\System (our)} & \bf \tab{\\Style} & \tab{\bf Single\\\bf Ch.} & \bf \tab{\\cpWER} %
    && \tab{\bf SAg-WER\\\bf DI-cpWER} \\
    \midrule
    TS-VAD \arrow GSS \arrow Base & D \arrow S \arrow A & \xmark & 6.70 && 4.36 \\  %
    TS-VAD \arrow WavLM & D \arrow A & \cmark & 9.26 && 7.16 \\
    TS-VAD \arrow GSS \arrow WavLM & D \arrow S \arrow A & \xmark & 5.77 && 3.40 \\  %
    \midrule
    TS-SEP \arrow Mask. \arrow Base & D + S \arrow A & \cmark & 11.61 && 9.66 \\ %
    TS-SEP \arrow GSS \arrow Base & D + S \arrow A & \xmark & 6.42 && 4.34 \\
    TS-SEP \arrow Mask. \arrow WavLM & D + S \arrow A & \cmark & 7.81 && 5.80 \\
    TS-SEP \arrow GSS \arrow WavLM & D + S \arrow A & \xmark & 5.36 && 3.27 \\
    \bottomrule
  \end{tabular}
\end{table}

\begin{table}
\centering

    \makeatletter \DeclareRobustCommand{\arrow}{%
\check@mathfonts%
\raisebox{\fontdimen22\textfont2}{\tikz{\draw[arrow](0,0)--(0.8em,0);}}%
\xspace}\makeatother

    \caption{\diff{}{Breakdown of \gls{cpWER} for different overlap subsets, from no overlap with long (0L) or short (0S) pauses to overlap ratios between 10 and 40 percent.}}
 \label{tab:res:breakdown}
  \setlength{\tabcolsep}{1.5pt}
  \begin{tabular}{lSSSSSSS}
    \toprule
    \tblrowadded & \multicolumn{6}{c}{Overlap ratio in \%} & \\
    \cmidrule{2-7}
    \tblrowadded\bf System & {0L} & {0S} & {10} & {20} & {30} & {40} & {avg} \\
    \midrule
    \tblrowadded TS-VAD \arrow GSS \arrow Base & 5.76 &  4.34 &  6.73 &  8.12 &  8.04 &  6.53 &  6.70 \\  %
    \tblrowadded TS-VAD \arrow WavLM & 5.05 &  5.04 &  7.89 & 10.78 & 12.27 & 11.90 &  9.26 \\  %
    \tblrowadded TS-VAD \arrow GSS \arrow WavLM & 4.66 & 3.51 &  5.96 &  7.59 &  7.02 &  5.24 & 5.77 \\ %
    \midrule
    \tblrowadded TS-SEP \arrow Mask. \arrow Base & 7.91 &  5.57 &  9.34 & 13.27 & 15.09 & 15.58 & 11.61 \\ %
    \tblrowadded TS-SEP \arrow GSS \arrow Base & 6.15 & 3.71 &  6.32 &  7.90 &  7.12 &  6.80 &  6.42 \\ %
    \tblrowadded TS-SEP \arrow Mask. \arrow WavLM & 5.25 & 3.63 &  6.63 &  9.53 &  9.84 & 10.10 &  7.81 \\ %
    \tblrowadded TS-SEP \arrow GSS \arrow WavLM & 5.02 & 2.75 &  5.35 &  7.14 &  5.91 &  5.55 & 5.36 \\ %
    \bottomrule
  \end{tabular}

\end{table}

\subsection{Comparison with the literature}

LibriCSS is a publicly available dataset and thus allows for a comparison of our system with the performance achieved by others. We compiled some relevant results in \cref{tab:res:final}. \diff{}{For completeness, we furthermore report the  performance of our systems as a function of the speech overlap condition  in \mbox{\cref{tab:res:breakdown}}}.

In our experiments until here, we followed the convention proposed in \cite{Raj2021Meeting} that the first session of LibriCSS is used for DEV and the remaining 9 for EVAL.
For some of the results from the literature that we report here, it is not clear from the publication if this convention is followed, or if the results are for DEV+EVAL.
For the sake of fairness, we thus report in this table our results for DEV+EVAL, as they are in our case slightly worse than for EVAL alone.

The best \glspl{WER} for LibriCSS that we found are \SI{3.34}{\percent} from Raj et al.~\cite{Raj2022gpuGSS} and \SI{3.97}{\percent} from Wang et al.~\cite{Wang2021CSSSpectralMapping}.
But both used oracle diarization to obtain those numbers.
Besides those experiments, Wang et al.~\cite{Wang2021CSSSpectralMapping} also reported, as far as we know, the best asclite-based \gls{WER} on LibriCSS of \SI{5.85}{\percent} by using multiple channels.
They used a \gls{CSS} pipeline, where a system produced two streams from the observation, each stream containing no overlap.
When they restricted the system to single-channel data, they obtained \SI{10.15}{\percent} \gls{WER}.
In a side experiment of \cite{Kanda22tSOT}\footnote{\diff{}{Visible in appendix of arXiv version at \mbox{\url{https://arxiv.org/abs/2202.00842}}}}, Kanda et al.{} reported an asclite-based \gls{WER} of \SI{7.6}{\%}, only utilizing a single channel for their serialized output training based system, which yields directly the transcriptions, instead of producing an intermediate overlap-free audio signal.
Later, Kanda et al.\ \cite{Kanda2022T2D} proposed Transcribe-to-Diarize, which includes diarization information, and reported a \gls{cpWER} of \SI{11.6}{\percent},
which is the best single-channel \gls{cpWER} that we found.

In \cite{Delcroix2021MeetinSpeakerBeam}, Delcroix et al.\ reported a combination of \gls{TSVAD} and Speakerbeam for single-channel processing and obtained a \gls{cpWER} of \SI{18.8}{\percent}.
To interpret this \gls{WER}, they also tried \gls{TSVAD} with \gls{GSS} and obtained a \gls{cpWER} of \SI{11.2}{\percent}, which is the best \gls{cpWER} (without single-channel constraint) that we found for LibriCSS.

In \cite{boeddeker22_interspeech}, an \gls{SMM} was applied to LibriCSS, but it had issues with the memory consumption and was therefore only applied to the subsegmented data with an average duration of around \SI{1}{min}, which is  typically used for \gls{CSS} pipelines with asclite-based \gls{WER} evaluations.
The \gls{SMM}
obtained a \gls{cpWER} of \SI{5.9}{\%} on the subsegmented data.
In \cite{Wang2022DOACSS}, a \gls{DOA} approach is used to do diarization after a \gls{CSS}-based separation.
This is one of the few publications that report \gls{cpWER} on both the \SI{1}{min} data and the full \SI{10}{minute} long mini sessions, and thus can be used to estimate the impact of the session length on the WER performance.
They observed an improvement of \SI{0.58}{\%} when using the shorter data.
An improvement is reasonable, since shorter data can compensate some diarization errors in \gls{cpWER}.

By using the publicly available pretrained ``base'' \gls{ASR} system from mid 2020, which was available for all mentioned references \diff{}{and used in\mbox{\cite{boeddeker22_interspeech}}}, we are able to outperform or be similar to all systems, except \cite{Kanda22tSOT} for the asclite-based \gls{WER} under the single-channel constraint.
By using the recently released ``WavLM'' based \gls{ASR}, we are able to outperform all systems.
Note that the combination of our reimplemented \gls{TSVAD} and the pretrained ``WavLM'', without any enhancements, is a single-channel system and has a better \gls{cpWER} on the 10-minute data than all references, regardless of whether they use single- or multi-channel data.

\diff{}{The diarization estimates of\mbox{\cite{Raj2022gpuGSS}} are publicly available}\footnote{\diff{}{\mbox{\url{https://github.com/desh2608/diarizer}}}}\diff{}{.
We applied \gls{GSS} and the ``WavLM'' based \gls{ASR} to them and observed a degradation of the \gls{cpWER}, which highlights the strength of our diarization performance.}

Interestingly, we also obtained some \glspl{WER} that are better than those from the literature while solving a more difficult task.
Without using oracle diarization, we were able to obtain a speaker agnostic \gls{WER} of \SI{3.27}{\%}, while the so far best \gls{WER} is \SI{3.34}{\%}~\cite{Raj2022gpuGSS}, where oracle diarization was used.
While \cite{Wang2021CSSSpectralMapping} uses a speaker agnostic \gls{WER} to obtain \SI{5.85}{\percent}, our proposed system is able to achieve a \gls{WER} of \SI{5.36}{\percent} while additionally counting errors caused by diarization.

\section{Conclusion \diffsimple{}{and outlook}}
\label{sec:conclusion}
By expanding the output from time to time-frequency resolution, we were able to extend the \gls{TSVAD} diarization system to a joint diarization and separation method.
The approach, which we named \gls{TSSEP}, is general in the sense that it can be used both for single- and multi-channel input data.
While \gls{TSSEP} can be viewed as an alternative to \gls{GSS}, it can also be combined with it for further performance improvement on multi-channel data.
We discussed various training details of \gls{TSVAD} and \gls{TSSEP}, and presented an extensive experimental evaluation on the LibriCSS meeting data set.
Performance was measured in terms of diarization error rate and in terms of both diarization dependent and independent word error rates, to assess the impact of diarization errors on ASR.
Obtaining \SI{11.61}{\%} and \SI{6.42}{\%} for the single- and multi-channel cases with a ``base'' ASR model, and \SI{7.81}{\%} and \SI{5.36}{\%} with WavLM features for ASR, we were able to achieve a new state of the art in diarization dependent WER.
The best result translates into a WER of \SI{3.27}{\%} for diarization error independent evaluation. 

\diff{}{Going forward, we plan to apply \gls{TSSEP} to real recordings such as CHiME-6.
Since \gls{TSSEP} can also be trained with the \gls{TSVAD} objective, a hybrid training is possible, where we simultaneously train with the \gls{VAD} objective on real data and with a separation objective on simulated data, in order to enable training on real data for separation without a separation target.}

\section*{Acknowledgments}
Part of this work was done while Christoph Boeddeker was an intern at Mitsubishi Electric Research Laboratories (MERL).
He was later funded by Deutsche Forschungsgemeinschaft
(DFG), project no. 448568305.
Computational resources were provided by the Paderborn Center for Parallel Computing.

\bibliographystyle{IEEEtran}

\bibliography{mybib}

\end{document}